\documentclass[preprint2]{aastex}

\def\cf{{c.f.,~}}

\newcommand{\tur}{4C~41.17}
\newcommand{\squ}{4C~60.07}
\newcommand{\hor}{B2~0902+34}

\newcommand{\etal}{{et al.}}
\newcommand{\eg}{{e.g.,}}
\newcommand{\lya}{\ifmmode {\rm Ly\alpha}\else{\rm Ly$\alpha$}\fi}
\newcommand{\hzrgs}{HzRGs}
\newcommand{\hzrg}{HzRG}

\newcommand{\LFIR}{\ifmmode {\rm \,L_{FIR}}\else ${\rm \,L_{FIR}}$\fi}
\newcommand{\LUV}{\ifmmode {\rm \,L_{UV}}\else ${\rm \,L_{UV}}$\fi}
\newcommand{\Llya}{\ifmmode {\rm \,L_{Ly\alpha}}\else ${\rm \,L_{Ly\alpha}}$\fi}
\newcommand{\OmM}{\ifmmode {\Omega_{\rm M}}\else $\Omega_{\rm M}$\fi}
\newcommand{\OmL}{\ifmmode {\Omega_{\Lambda}}\else $\Omega_{\Lambda}$\fi}
\newcommand{\ph}{\ifmmode {h_{71}^{-1}}\else $h_{71}^{-1}$\fi}
\newcommand{\psqh}{\ifmmode {h_{71}^{-2}}\else $h_{71}^{-2}$\fi}

\def\HII{{\ion{H}{2}}}

\newcommand{\kms}{\,\mbox{km s$^{-1}$}}
\newcommand{\degree}{\ifmmode {^{\,\circ}} \else {$^{\,\circ}$}\fi}
\newcommand{\nh}{\ifmmode {N_{\rm H}} \else {${N_{\rm H}}$}\fi}
\newcommand{\Lsun}{\ifmmode {\rm\,L_\odot}\else ${\rm\,L_\odot}$\fi}
\newcommand{\Msun}{\ifmmode {\rm\,M_\odot} \else ${\rm\,M_\odot}$\fi}
\newcommand{\Zsun}{\ifmmode {\rm\,Z_\odot} \else ${\rm\,Z_\odot}$\fi}
\newcommand{\kmps}{\ifmmode {\rm\,km~s^{-1}} \else ${\rm\,km\,s^{-1}}$\fi}
\newcommand{\kpc}{{\rm\,kpc}} 
\newcommand{\ergps}{\ifmmode {\rm\,erg\,s^{-1}} \else {${\rm\,erg\,s^{-1}}$}\fi}
\newcommand{\ergpspcm}{\ifmmode {\rm\,erg\,s^{-1}\,cm^{-2}} \else {${\rm\,erg\,s^{-1}\,cm^{-2}}$}\fi}
\newcommand{\surfbr}{\ifmmode {\rm\,erg\,s^{-1}\,cm^{-2}\,arcsec^{-2}} \else {${\rm\,erg\
,s^{-1}\,cm^{-2}\,arcsec^{-2}}$}\fi}
\newcommand{\Msunpyr}{\ifmmode {\rm\,M_\odot\,yr^{-1}} \else {${\rm\,M_\odot\,yr^{-1}}$}\fi}
\newcommand{\pyr}{\ifmmode {\rm\,yr^{-1}} \else {${\rm\,yr^{-1}}$}\fi}
\newcommand{\psec}{\ifmmode {\rm\,s^{-1}} \else {${\rm\,s^{-1}}$}\fi}

\def\kms{\ifmmode {\rm\,km\,s^{-1}}\else
     ${\rm\,km\,s^{-1}}$\fi}
\def\ergs{\ifmmode {\rm\,erg\,s^{-1}}\else
     ${\rm\,erg\,s^{-1}}$\fi}
\def\hone{\ion{H}{1}}

\def\heii{\ion{He}{2}}
\def\heiil{\ion{He}{2} $\lambda$1640}
\def\lya{Ly$\alpha$}

\def\nv{\ion{N}{5}}

\def\ciii{\ion{C}{3}]}

\def\civ{\ion{C}{4}}
\def\civl{\ion{C}{4} $\lambda$1549}

\def\oii{[\ion{O}{2}]}
\def\oiil{[\ion{O}{2}] $\lambda$3727}

\def\hbeta{H$\beta$}
\def\oiii{[\ion{O}{3}]}
\def\oiiil{[\ion{O}{3}] $\lambda$5007}

\received{}
\begin{document}

\title{Metal enriched gaseous halos around distant radio galaxies:\\
Clues to feedback in galaxy formation \thanks{Based on observations at
the W.M. Keck Observatory, which is operated as a scientific
partnership among the University of California, the California
Institute of Technology, and the National Aeronautics and Space
Administration.  The Observatory was made possible by the generous
financial support of the W.M. Keck Foundation.}}

\author{Michiel Reuland\altaffilmark{1,2,3}, Wil van
Breugel\altaffilmark{1,4}, Wim de Vries\altaffilmark{1,2}, Michael
A. Dopita\altaffilmark{5}, Arjun Dey\altaffilmark{6}, George
Miley\altaffilmark{3}, Huub R\"ottgering\altaffilmark{3}, Bram
Venemans\altaffilmark{3}, S.A. Stanford\altaffilmark{1,2}, Mark
Lacy\altaffilmark{7}, Hy Spinrad\altaffilmark{8}, Steve
Dawson\altaffilmark{8}, Daniel Stern\altaffilmark{9} \& Andrew
Bunker\altaffilmark{10} }

\altaffiltext{1}{Institute of Geophysics and Planetary Physics,
Lawrence Livermore National Laboratory, L-413, Livermore, CA 94550
USA}
\altaffiltext{2}{Physics Department, University of California at
Davis, One Shields Avenue, Davis, CA 95616 USA}
\altaffiltext{3}{Sterrewacht Leiden, Postbus 9513, 2300 RA Leiden The
Netherlands} 
\altaffiltext{4}{University of California, Merced, P.O. Box 2039,
Merced, CA 95344} 
\altaffiltext{5}{Research School of Astronomy and Astrophysics, The
Australian National University, Cotter Road, Weston Creek, ACT2611,
Australia} 
\altaffiltext{6}{NOAO, 950 N. Cherry Ave., Tucson, AZ 85719 USA}
\altaffiltext{7}{SIRTF Science Center, Caltech, MS 220-6, 1200
E. California Boulevard, Pasadena, CA 91125 USA}
\altaffiltext{8}{Department of Astronomy, University of California at
Berkeley, Berkeley, CA 94720 USA} 
\altaffiltext{9}{Jet Propulsion Laboratory, California Institute of
Technology, Mail Stop 169-327, Pasadena, CA 91109 USA}
\altaffiltext{10}{Institute of Astronomy, University of Cambridge,
Madingley Road, Cambridge CB3 0HA UK}

\email{wil@igpp.ucllnl.org}

\begin{abstract}

We present the results of an optical and near-IR spectroscopic study
of giant nebular emission line halos associated with three $z > 3$
radio galaxies, \tur, \squ\ and \hor.  Previous deep narrow band \lya\
imaging had revealed complex morphologies with sizes up to 100\kpc),
possibly connected to outflows and AGN feedback from the central
regions. The outer regions of these halos show quiet kinematics with
typical velocity dispersions of a few hundred \kmps, and velocity
shears that can mostly be interpreted as being due to rotation. The
inner regions show shocked cocoons of gas closely associated with the
radio lobes. These display disturbed kinematics and have expansion
velocities and/or velocity dispersions $>$1000\kmps\ . The core region
is chemically evolved, and we also find spectroscopic evidence for the
ejection of enriched material in \tur\ up to a distance of
$\approx$60\kpc\ along the radio-axis. The dynamical structures traced
in the \lya\ line are, in most cases, closely echoed in the Carbon and
Oxygen lines. This shows that the \lya\ line is produced in a highly
clumped medium of small filling factor, and can therefore be used as a
tracer of the dynamics of \hzrgs. We conclude that these \hzrgs\ are
undergoing a final jet-induced phase of star formation with ejection
of most of their interstellar medium before becoming ``red and dead"
Elliptical galaxies.

\end{abstract}


\section{Introduction}

There is compelling evidence that in galaxies the formation and
evolution of the central stellar bulge and the massive nuclear black
hole are intimately related
\citep{Magorrian98,FerrareseMerritt00,Gebhardt00}. Understanding this
coevolution of galaxy spheroids and their central massive black holes
is one of the major outstanding issues in modern cosmology. Because
they are readily located by their ultra-steep spectrum radio
properties, the high redshift radio galaxies (\hzrgs; $z > 3$) provide
an efficient means to locate and study the environment and physics of
newly-forming galaxies. There are at least two reasons why \hzrgs\ are
key in attempts to understand the physical processes involved. First,
they rank among the most luminous, largest, and most massive galaxies
known in the early Universe \citep[\eg][]{DeBreuck02}. Secondly, we
observe them early in the epoch of galaxy formation at a time when
their super-massive black holes (SMBHs) are highly active, and while
their relativistic jets are interacting most strongly on their host
galaxies.

Within the framework of the standard $\Lambda$ Cold Dark Matter
($\Lambda$-CDM) scenario, it is believed that massive galaxies grow in
a hierarchical fashion through the merging of smaller stellar and dark
matter halo objects. Whether their central black holes grow in similar
fashion or whether they are primordial objects
\citep{Loeb93,SilkRees98,KauffmannHaehnelt00}, the energetic outflows
and the ionizing radiation from these central active SMBHs are
expected to profoundly influence the evolution both of their parent
galaxies and of the surrounding environment. Recent models describing
the formation of massive galaxies and clusters provide further
evidence for the importance of this feedback
\citep{Benson03,Springel05}. Additionally, outflows could provide a
source for the chemical enrichment of the inter galactic medium seen
at high redshifts (e.g., Rauch, Sargent, \& Barlow 2001; Aguirre et
al. 2001).  While there is consensus that feedback in some form must
be important, there is an ongoing debate about which manifestations
dominate and the scales on which they operate.

Because they are themselves massive galaxies in the early universe,
\hzrgs\ are located in the regions of large matter overdensities and
within regions rich in forming galaxies.  Indeed, many are known to be
embedded in regions of dense interstellar medium (ISM) and in
environments containing the earliest known galactic clusters
\citep[\eg][]{McCarthy93,vanOjik96,vanOjik97,Athreya98,Pentericci00a,Papadopoulos00}.
These gaseous reservoirs enable us to study in detail the feedback
processes occurring at high redshift.

In an earlier paper \citep{Reuland03} we described the results from
narrow-band imaging observations of unprecedented sensitivity of three
\hzrgs\ (\hor, \squ, and \tur). These images were obtained at the Keck
II 10\,m telescope using custom-made, high-throughput interference
filters with bandpasses centered at the redshifted \lya\ line. The
observations revealed very luminous ($L_{ \rm Ly\alpha} \approxeq
10^{45} \ergs$) and extended ($\approx$200\,\kpc) emission line
nebulae with spectacular features not previously seen, such as long
filamentary structures, ionization cones, and multiple sharply bounded
regions of enhanced emission, all indicative of strong interactions as
expected from the scenario painted above.  We argued that these
extended \lya\ nebulae might represent gas cooling in massive CDM
halos, supplying new material for the continued growth of the galaxies
at their center. This feeding could then be responsible for feedback
mechanisms through radio jets, supernova explosions, and radiation
pressure from the AGN, resulting in large scale outflows.  The
extended X-ray halo around \tur\ provides further evidence for a
highly interactive environment in such systems \citep{Scharf03}.

However, with only morphological data, important issues concerning
possible origins for the filaments and large-scale structures well
beyond the radio sources could not be tackled.  Because \lya\ is a
resonance line, it is also important to resolve the question of
whether the nebulae represent truly extended ionized halos, or rather,
are due to \lya\ photons which are produced near the central AGN and
are scattered off neutral hydrogen halos.  Many other questions
concerning the structure, origin, and fate of the emission-line gas
remain. For example: is the gas organized in shells, filaments, or
cloudlets? What is the source of ionization and what is the chemical
composition of the gas? Can the outflows regulate the growth of
galactic bulge and black-hole? Can they expel metals from the deep
potential wells to enrich the intergalactic and inter-cluster medium?

In order to better understand the kinematics, abundances, and
ionization mechanisms of these halos we obtained optical and
near-infrared spectra with the Keck telescopes to measure the extent,
intensity, and kinematics of \lya\ and other \emph{non-resonance}
lines, [{\ion{He}{2}}], [\ion{O}{2}] and [\ion{O}{3}], at various
position angles across the nebulae.  This paper discusses the results
of these spectroscopic observations.

The structure of the paper is as follows: The sample selection,
observations and data analysis are described in \S 2.  In \S 3 we
present results for individual objects. \S 4 is a discussion of the
observed kinematics and line ratios of the halos.  The implications
for the origin and fate of the emission line halos are discussed in \S
5 and our conclusions are given in \S 6.

Here we adopt the concordance cosmological parameters $\OmM = 0.27$,
$\OmL = 0.73$ and $H_{0} = 71 \kmps\,\rm Mpc^{-1}$. The age of the
Universe is $1.7-1.9$\ph\,Gyr at the redshifts ($z = 3.4 - 3.8$) of
our galaxies, and the angular-to-linear transformation is
$\approx$7.4\ph\kpc\ arcsec$^{-1}$.

\section{Observations and data analysis}

\subsection{Sample Selection}

The three objects selected for the spectroscopic follow-up
observations are shown in Table \ref{Table1} with their positions and
adopted redshifts. They were selected from amongst the galaxies
observed in the course of our Keck imaging program
\citep{Reuland03}. The reasons for their inclusion in the
spectroscopic program are summarized as follows.

\tur\, at $z = 3.8$, was one of the first \hzrgs\ to be discovered
\citep{Chambers90} and for many purposes serves as an archetype
\hzrg. Optical \citep{Dey97} as well as sub-mm wavelength observations
\citep{Dunlop94, Ivison2000} have shown that it is a massive forming
galaxy with a star formation rate of up to several thousand \Msunpyr.
Recently, very extended X-ray emission was found around \tur, which
follows the \lya\ morphology closely \citep{Scharf03}.

We selected \squ\ \citep[$z=3.8$;][]{Chambers96,Rottgering97} because
it shows both spatially and kinematically resolved CO emission
\citep{Papadopoulos00,Greve05}. Interestingly, in \citet{Reuland03} it
was found that the \lya\ halo has a very extended ($76$\ph\kpc)
filament which appears orthogonal to the major axis of the CO and dust
emission.

\hor\ \citep[$z=3.4$;][]{Lilly88} was selected because it is thought
to be a protogalaxy, dominated by young stars
\citep{EisenhardtDickinson92}. So far, it is one of only a handful of
\hzrgs\ for which neutral hydrogen has been detected in absorption
against the radio continuum \citep{Uson91,CodyBraun03}.

\subsection{Optical and Near-Infrared Spectroscopy}

\begin{figure*}[t]
\epsscale{2.0}
\plotone{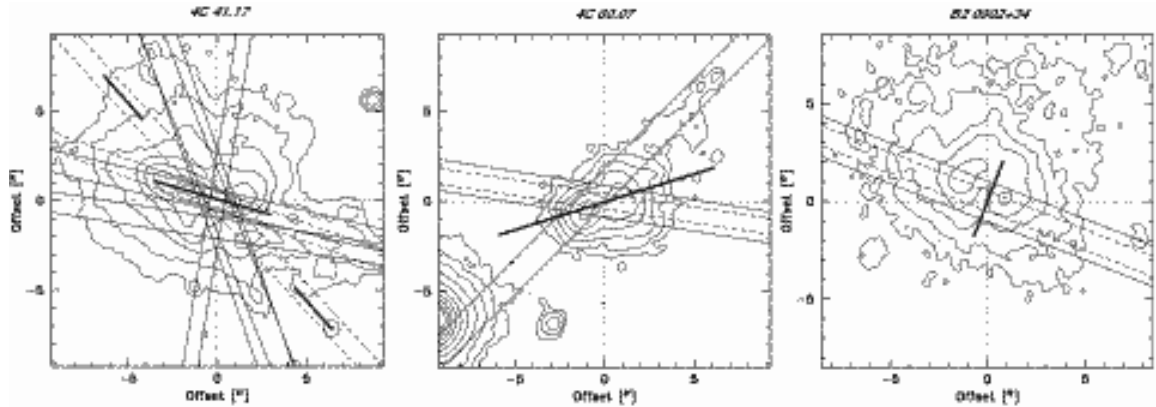}
\caption{\label{fig1} Contour representations of the \lya\ emission
line halos around \tur, \squ, and \hor\ (left, middle, right) with the
various PAs at which they were studied overlaid. In this figure, north
is at the top and east at the left. The solid and dashed lines
correspond with slit positions of the optical and near-IR spectroscopy
respectively and the dotted lines indicate the positions of the radio
cores. The bold lines represent the direction of the radio axes. In
the case of \tur\ two axes are shown, corresponding to the inner and
outer radio lobes. For each object the contours indicate observed
surface brightnesses of
$6.7\times10^{-19}\times(6,12,25,50,100,200,400,800)$
erg~s$^{-1}$~cm$^{-2}$~arcsec$^{-2}$; see \citet{Reuland03}.}
\end{figure*}

Figure \ref{fig1} shows the different slit positions used in the
program overlaid on contour representations of the narrow-band \lya\
images of the galaxies. In contrast to most previous spectroscopic
studies of \hzrgs, most of the slits were not placed directly along
the radio axis. The specific instrumental setups are given in Table
\ref{Table2}. The data reduction techniques are described below.

\subsubsection{Optical Observations: LRIS}

Most of the optical observations were carried out using the
Low-Resolution Imaging Spectrometer \citep[LRIS;][]{Oke95} at the
Cassegrain focus on the 10\,m Keck I Telescope. The data were
collected with various instrumental setups using both the long-slit
mode and multi-slit masks designed to obtain spectra for $\approx 15$
targets in the field simultaneously (as part of a survey looking for
associated galaxies in the proto-cluser; Croft \etal\ 2006 in
preparation). The red-sensitive LRIS-R camera was employed. This uses
a Tektronix 2048$\times$2048 CCD detector with a pixel scale of
$0\farcs215$\,pixel$^{-1}$.

All of the spectroscopic reductions were performed using standard
methods and the NOAO IRAF\footnote{IRAF is distributed by the National
Optical Astronomy Observatory, which is operated by the Association of
Universities for Research in Astronomy, Inc., under cooperative
agreement with the National Science Foundation.} package
\citep{Tody93}. Skylines were used to improve the first order
wavelength calibration based on arc spectra to better than 0.3\,\AA\
rms. The instrumentral resolution was measured from the unblended
skylines. Flux calibrations were performed using observations of
standard stars such as Feige 110 and Feige 34 \citep{Massey88}. The
extended emission of \tur\ filled the narrow slits of the multi-object
spectroscopic program, rendering accurate sky subtraction
difficult. This does not seriously affect the kinematics and relative
fluxes of interest for this paper.

\subsubsection{Optical Observations: ESI}

One set of observations along the filament of \squ\ was made during
the night of UT 2001 February 25, using the Echelle Spectrograph and
Imager \citep[ESI;][]{Sheinis00} at the Cassegrain focus of the Keck
II 10\,m telescope in low-dispersion mode.  The detector used is a
high-resistivity MIT-Lincoln Labs 2048 $\times$ 4096 CCD with a plate
scale of 0\farcs154 pixel$^{-1}$.  Exposures were broken into
integrations of 1800 seconds each, one of which had to be halved
because of time constraints. We performed 6\arcsec\ offsets between
each integration. The data were rotated over an angle depending on the
position of the object on the slit, to align the dispersion axes. The
wavelength calibration varies slightly with slit position, hence we
shifted the spectra along the dispersion axes.  This first order
approximation is sufficiently accurate for the region of interest
4300\AA$-$9200\AA\ (corresponding to approximately 900\AA$-$1920\AA\
in the rest frame). Subsequent data reduction was done using standard
methods in IRAF.


\subsubsection{Near-IR Spectroscopy: NIRSPEC}

The near-infrared spectra were obtained using the 10\,m Keck II
Telescope with its Near-Infrared Spectrograph
\citep[NIRSPEC;][]{McLean98}.  The slit dimensions were 0.76\arcsec\
$\times$ 42\arcsec\ slits giving low-resolution (R $\approx$
1400$-$1900) spectra in wavelength ranges chosen to include the
\hbeta, \oii\ and \oiii\ lines of the target galaxies (see Table
\ref{Table2} for details).  In this low-resolution mode, the
1024$\times$1024 ALADDIN InSb detector has a plate scale of
0\farcs143\,pixel$^{-1}$. We obtained sets of 900\,s integrations each
with $\approx 5-10$\arcsec\ spatial offsets between exposures.

The NIRSPEC spectra need to be corrected for the spectral curvature
and spatial distortions caused by the high-throughput optics. A
general correction would require rectification onto a slit
position-wavelength grid based on a wavelength solution from skylines
and coadded exposures of a standard star. However, since no continuum
is apparent in our spectra, we have only extracted small regions lying
close to the emission lines. This approach requires only a simple
rotation over an angle which depends on the wavelength of interest to
provide a local calibration of the wavelength along the slit.

The data were flat-fielded and corrected for cosmic rays and bad
pixels in the standard fashion. In order to remove the strong near-IR
skylines, and a sky frame scaled to the brightness of unsaturated sky
lines near the emission line of interest was subtracted.
Subsequently, the frames were cropped, rotated, and coadded. Flux
calibration was done with standard stars of spectral type A0V, B3, and
G4, and was consistent to within 10\%.

\subsection{Data analysis}

For the data analysis the spectra were registered in position with
radio maps from the literature \citep{Carilli94,Carilli95,Carilli97}.
The zero points of the spatial scales correspond with the radio
core. This was achieved by identifying the core with the centroid of
the continuum emission. If no continuum emission was visible, we
identify the core with either the spatial region that shows the
broadest line emission, or by bootstrapping to spectra for which the
core could be reliably identified. These results were then checked
with the narrow-band images, correlating with the peaks and dips in
the observed intensity. This resulted in a total uncertainty less than
0.5\,\arcsec\ in the relative spatial offsets.

The velocity scale used in the analysis is relative to the systemic
velocity derived from the \heiil\ line (the adopted systemic redshifts
are given in Table \ref{Table1}). The kinematic information was
obtained from the 2-D spectra using a program written in IDL making
use of the publicly available fitting routine MPFIT\footnote{Available
at http://cow.physics.wisc.edu/$\approx$craigm/idl/fit\-ting.html}.
The data were coadded within apertures matched to the seeing in order
to increase the signal-to-noise ratio and ensure that the extracted
spectra are not correlated. We then determined the velocity centroid,
FWHM (corrected for instrumental broadening measured from unsaturated
skylines, and assuming that the widths add in quadrature). The peak
and baseline fluxes were determined by fitting a single Gaussian and
baseline to each trace. Single Gaussians provide a good fit to the
outer regions and in order to allow a direct comparison with the
central regions similarly, we have chosen to treat all regions
consistently using the simpler approach. The single Gaussian
decomposition for high surface brightness emission may break down when
these regions are embedded in more smoothly varying large scale low
surface brightness envelopes.

\section{Dynamical Results}

\begin{figure}[t]
\epsscale{1.0}
\plotone{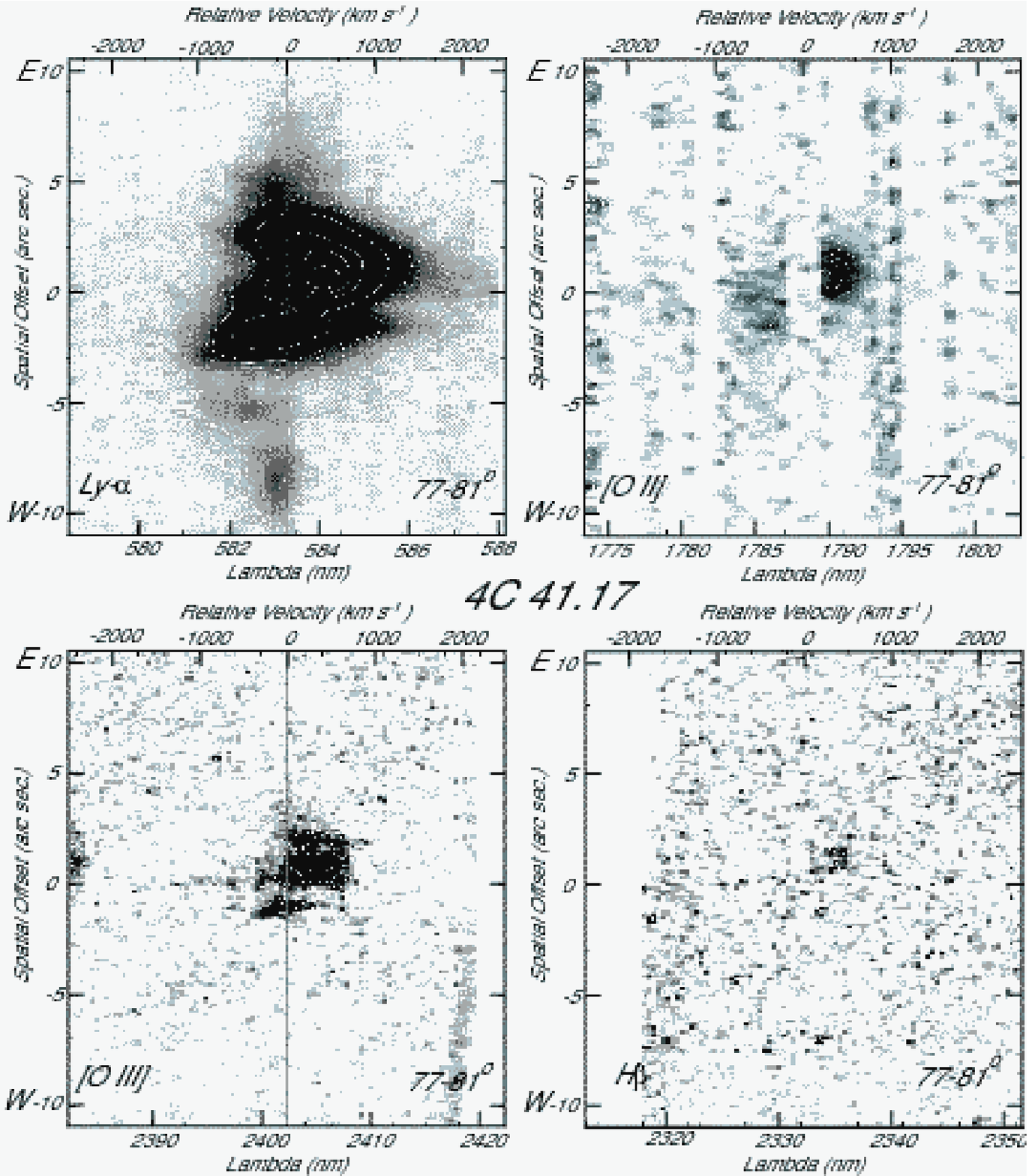}
\caption{\label{fig2} Grayscale representations of the 2-D spectra of
\tur\ centered at the \lya, \oiil, \hbeta, and \oiiil\ lines. The
velocities are indicated relative to the systemic redshift determined
from the \heii\ line. The zero points of the spatial scales correspond
with the position of the radio core as identified with the broad line
and continuum emission. Note that the peaks of the line emission
corresponding with the position of radio knot B2 in \citet{Carilli94}
are 1\,\arcsec\ E of the nucleus. It can be seen (most easily in the
\oiiil\ spectrum) that there are 0.5\arcsec\ offsets between the broad
lines and the central depression. The \oii\ spectrum is rich in very
strong skylines. These have been masked off to better show the
structure of the \oii\ line emission.}
\end{figure}

Figures \ref{fig2}, \ref{fig5}, and \ref{fig9} show 2-D optical and
near-IR spectra of \tur, \squ, and \hor\ centered at the emission
lines most relevant to our discussion.  It is immediately obvious,
that many of the lines are very broad, and show strong spatial
variation both in their velocity centroids and their FWHMs.  In some
cases continuum emission was also detected, but only from the central
regions, within 3\,\arcsec of the nucleus. Since the main interest of
this paper is the extended emission, here we will focus our discussion
on the nebular lines.

For both \tur\ and \squ\ the oxygen line profiles closely resemble the
bright inner parts of the \lya\ emission. Since self-absorption cannot
be important in the strong forbidden lines, this suggests that the
\lya\ line is giving useful information about the kinematic structure
despite being subject to resonance scattering. This can only be
possible if the interstellar gas is highly clumped. If this is true in
the inner regions, it is probably also true in the outer regions. In
any case, even if the line is being resonantly scattered in the outer
regions, it would likely still provide a fair (although possibly
velocity biased) measure of the neutral gas kinematics.

\begin{figure}[t]
\epsscale{1.1}
\plotone{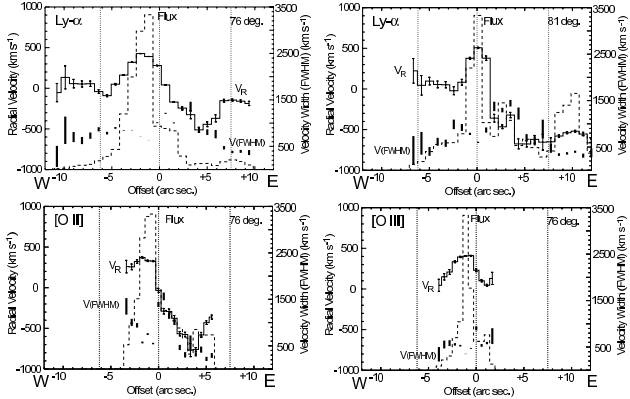}
\caption{\label{fig3} Relative velocities (solid lines), velocity
dispersion (bars), and normalized surface brightness profiles (dashed
lines) as determined from spectra with slit positions parallel to the
radio axis for the \lya, \oii, and \oiii\ emission lines of \tur.  Top
to bottom: \lya\ along PA=76\degree, \lya\ for PA=81\degree, \oii\ for
PA=70\degree, and \oiii\ for PA=68\degree. The spatial zeropoints
correspond with the position of the radio core and the dotted lines
represent the position of the radio lobes projected on the slit. East
is at the right in all of these panels. The bar and symbol size
indicate the 1$\sigma$ uncertainties on the measurements. Note that
the fits to the \oii\ emission are affected by the strong skylines and
that the slit along PA=81\degree\ did not go through the radio core.}
\end{figure}

\begin{figure}[t]
\epsscale{1.0}
\plotone{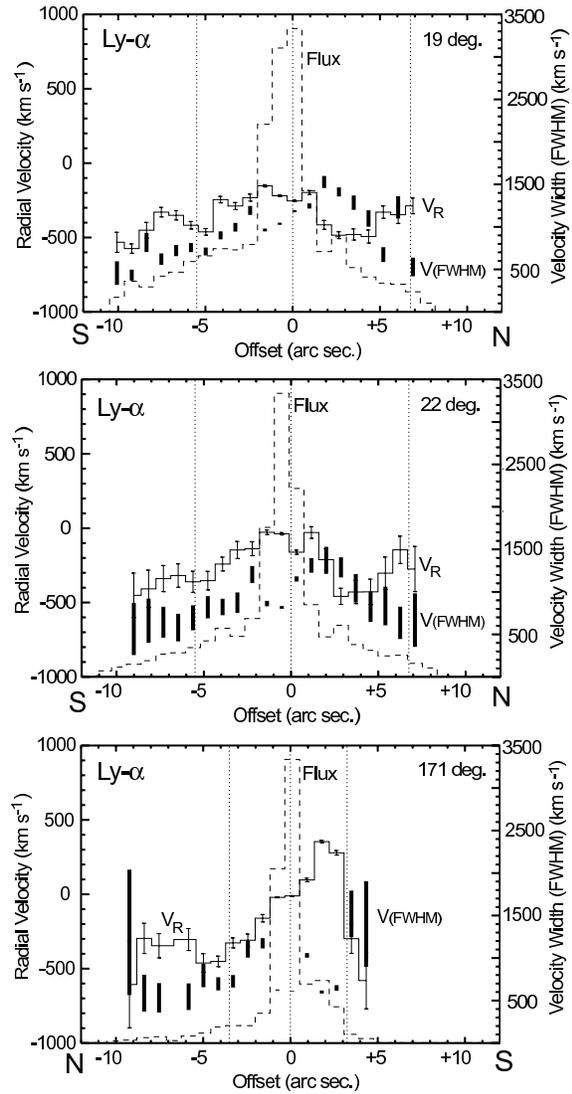}
\caption{\label{fig4} Similar to Fig. \ref{fig3} but for the \lya\
emission along the slits that lie more or less perpendicular to the
radio axis of \tur.  PA=19\degree\ (top), PA=22\degree\ (middle), and
PA=171\degree\ (bottom).}
\end{figure}

Figures \ref{fig3}, \ref{fig4}, \ref{fig6}, and \ref{fig10}, summarize
radial velocities, FWHMs, and the run of relative surface brightness
of the halos as a function of distance from the radio core for
different position angles. These figures show clear evidence for a
distinction between disturbed and more quiescent regions.  The inner
regions have higher surface brightnesses, are characterized by large
velocity dispersions (FWHM $\approx 1500$\,\kmps), and seem to be
embedded in low surface brightness regions with FWHMs of order
$\approx 500$\,\kmps. In \S \ref{Discussion} we will discuss the
general implications of these diagrams. First, we describe the
individual sources.

\subsection{Notes on individual objects}

\subsubsection{\tur: General Remarks} \label{4c41results}

\citet{Dey97} have discussed the brightest part ($2\arcsec \times
1\arcsec$) of the extracted 1-D optical spectrum of \tur\ at PA =
76\degree\ in detail. They determined a redshift $z=3.79786 \pm
0.00024$ based on the \heii\ line.  Furthermore they found evidence
for stellar absorption lines, and low polarization indicating that a
young stellar population contributes significantly to the rest-frame
UV continuum emission.

The 2-D spectra along the radio axis (PA 76\degree\ plotted in
fig. \ref{fig3}) show that the \lya, \oii, and \oiii\ emission line
regions are very extended (over approximately 20\,\arcsec,
10\,\arcsec, and 6\,\arcsec, respectively).  The \lya\ and \oiii\
lines both show two separate components straddling the radio core with
peak fluxes separated by about 3\,\arcsec. Careful inspection reveals
that these components are present also in the \oii\ line, with the red
part of the western component missing due to a skyline near $\lambda =
17880$\,\AA. The dip in between the emission peaks is expected since
the narrow-band \lya\ and radio image overlays showed the core to be
highly obscured (see \citet{Reuland03}).

The large extent of the nebula in the \oii, and \oiii\ emission lines
shows immediately that the whole \lya,\ halo is enriched in heavy
elements. This immediately discounts the possibility that the emission
line halo represents in-falling pristine gas. This point will be
discussed further below.

A second item of note is that the velocity gradient in the halo and
the velocity dispersions observed in the lines are comparable. This
shows that the halo is a highly turbulent and disordered structure.

The image overlays indicated that the position of the radio core is
offset by 0.5\,\arcsec\ to the NE from the central dip in the
emission. In the spectra, the position of the radio core is revealed
by a narrow high velocity tail on the \oiii\ profile, extending out to
-1500 \kms\ and clearly visible in figure \ref{fig2}. In the \lya\
profile, the nucleus is marked by a faint tail extending out to over
+2000 \kms\, the negative velocity tail presumably being obscured by
resonant scattering.

\begin{figure}[t]
\epsscale{1.0}
\plotone{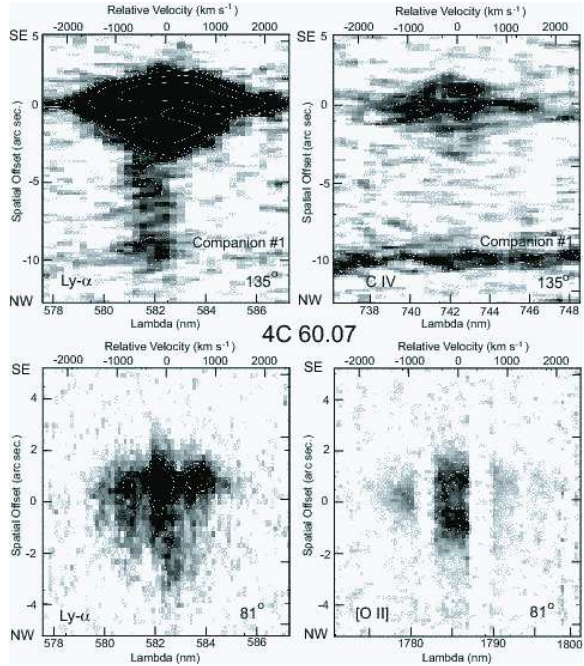}
\caption{\label{fig5} Similar to fig. \ref{fig2}. Grayscale
representations of the 2-D spectra of emission line halo around
\squ. The top left and top right panel respectively show \lya\ and
\civ\ emission along the filament with PA = 135\degree. The bottom
left and right panel show \lya\ and \oiil\ along the radio axis with
PA = 81\degree.}
\end{figure}

\subsubsection{\tur: Kinematics} 

\citet{Dey97} discerned the presence of both narrow and broad
components in the central regions of the \lya\ and \ciii\ emission
lines of \tur\ with FWHMs of $\approx$500-600\,\kmps\ and
$\approx$1200-1300\,\kmps\ respectively. Figure \ref{fig3} shows that
these high velocity dispersions extend along the radio axis to the
outer radio lobes. Beyond the limit of the radio lobes there is a
break in velocity, the kinematics become more quiescent, and the
velocity centroids change from being blue-shifted ($-$500 to
$-$700\,\kmps\ along the Western lobe) to near systemic. Perpendicular
to the radio axis (see Fig. \ref{fig4}) the kinematics follow a smooth
gradient and the broad emission lines are found only in the central
region of $\approx$3\arcsec\ ($\approx$20\,kpc) wide, identified by
its sharp peak in peak surface brightness. This bright emission is
associated with the inner radio lobes, and has been successully
ascribed to radiative shocks at the boundary of the expanding cocoon
\citep{Bicknell00}.

Except for the high velocity tail originating close to the AGN and
discussed above, the Oxygen lines show narrower line components than
the \lya\ line, with FWHMs $\approx$600\,\kmps. The measured
velocities and velocity dispersions likely reflect the true motions of
the gas, whereas \lya\ is additionally broadened by resonant
scattering into the damping wings.  The measurements on the \oiii\
line are likely to be much more reliable than the \oii\ line because
the line is brighter, is less reddened, and is not so badly cut about
by night sky lines. Furthermore, corrections to the FWHM do not need
to be made (in contradistinction to the \oii\ doublet), and finally,
it can be observed closer to the nuclear region than \oii, because of
its higher critical density \citep[see \eg\ ][]{diSeregoAlighieri97}.

Two very important results are found from the near-IR
spectroscopy. First, as has been noted in the previous section,
despite the obvious resonant broadening, the velocity structure of
\lya\ closely resembles that of \oiii. Secondly, there is \oii\
emission in the velocity regime $-$200 to $-$1000\,\kmps\ extending
from the nucleus to 7-8\arcsec\ west of the nucleus, whereas the
\oiii\ emission appears to be much more centrally concentrated. We
will discuss this further in \S \ref{Elines}.

The \lya\ emission appears to show a strong velocity shear in its
kinematically quiescent outer parts as yet undisturbed by the
expansion of the radio source along the radio axis. By contrast, the
kinematic profiles obtained with slit positions perpendicular to the
radio axis (see fig. \ref{fig4}) show a fairly symmetric velocity
distribution. This is indicative of an overall shear or rotation of
the $\approx 200$~kpc diameter \lya\ halo about its major axis.

The velocity structure of the bulk of the emission line gas appears
predominantly red-shifted in the Oxygen, \hbeta, and \lya\ lines. A
similar red-shift is seen in the extended CO J=4$-$3 emission
\citep{DeBreuck04}, suggesting that the AGN may be offset from the
systemic redshift of the galaxy. The central CO component is situated
at a relative velocity of $-$125\,\kmps. The location of this CO
component coincides in spatial position and velocity with the \lya\
gap between the ``cloud'' and the galaxy as discussed in
\citet{Reuland03}, suggesting that the dense molecular gas is
absorbing the \lya\ emission at this location.

\subsubsection{\squ: General Remarks}

The ESI spectrum of \squ\ (PA = 135\,\degree; fig. \ref{fig5}) shows a
clear trace of continuum emission over approximately 3\,\arcsec\ which
we associate with the position of the radio core and the unobscured
optical nucleus. Comparison of this identification with the extent of
the \lya\ filament extending towards to N-W narrow-band image
(Fig. \ref{fig1}) and which is visible in the lower part in the
spectrum, shows that the nucleus has been located to within
0.5\,\arcsec.  Based on the \heii\ line at 7855.7 $\pm$ 1.1\,\AA\ we
infer a redshift $z = 3.7887 \pm 0.0007$, in agreement with
\citet{Rottgering97}.

\subsubsection{\squ: Kinematics}

\begin{figure}[t]
\plotone{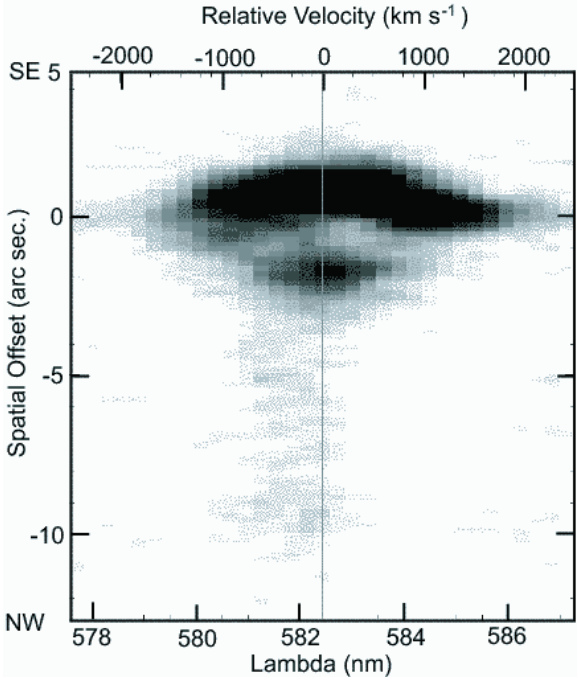}
\caption{\label{fig6} An expanded grayscale representation of the 2-D
spectrum of \squ\ centered at the redshifted \lya\ line showing the
filament and the crescent shaped arcs surrounding a depression in the
emission that is identified with the location of the both radio core
and unobscured optical nucleus. This kinematical signature is probably
the result of a shock interaction between the radio lobe and its
cocoon.}
\end{figure}

The 2-D \lya\ emission line (Fig. \ref{fig6}) shows interesting
structure: crescent shaped clouds surround a gap at the position of
the radio core near the systemic velocity of the galaxy. The spatial
structure was expected from the narrow-band image, but the kinematic
substructure is unusual. A comparable lack of \lya\ emission near the
systemic velocity of the galaxy has been found in a \lya\ galaxy
associated with the SSA-22 protocluster \citep{Wilman05} and the large
\lya\ halo recently discovered by the {\it Spitzer Space Telescope}
\citep{Dey05}.  It seems reasonable to attribute these velocity
profiles to complex radiative transport effects of the resonantly
scattered \lya\ emission in dense (dusty) media \citep[\eg\
][]{Neufeld90,Ahn04} since these sources are strong submillimeter
(rest-frame far-IR) emitters
\citep{Papadopoulos00,Chapman01}. However, the similarity of the
profiles of the the \lya\ line and the \civ\ line in this region
suggest, rather, an interaction of the radio lobe with a cocoon of
material.

The morphologies of the \lya\ and \oii\ spectra at PA = 81\degree\
(fig \ref{fig5}) are remarkably similar, providing further
justification for the use of \lya\ as a useful tracer of the dynamics
of these systems, despite the problems of resonance scattering Both
lines have very high velocity FWHMs of $\approx 100$\,\AA\ or $\approx
1600-1700$\,\kmps, rather large for \hzrgs. They are however smaller
than for the spectrum at PA = 135\degree\ which shows a FWHM of
$\approx 2600$\,\kmps\ . All of this is suggestive of the strong
radiative shocks produced in an interaction of a powerful radio source
with a surrounding dense interstellar medium.

The emission line profile across the \lya\ gap can be fit with two
Gaussian components with central wavelenghts of $5839.9 \pm 0.5$\,\AA\
and $5806.8 \pm 0.4$\,\AA\ and FWHMs of $12.8 \pm 0.5$\,\AA\ and $
10.6 \pm 0.3$\,\AA\ respectively.  However, it is probably better to
fit the whole bright region to a single expanding shell by a velocity
ellipsoid, estimated by fitting an ellipse to the position~:~velocity
curve obtained for the two peaks in line intensity at each spatial
position.  The fit obtained depends somewhat on both the position
angle of the observation and on the line which is used to measure it,
but all are consistent with an expansion velocity of $1100\pm
200$\,\kmps\ .

\subsubsection{\squ: The \lya\ Filament}

Perhaps the most striking feature of the emission line nebula
surrounding \squ\ is the extended \lya\ filament. Imaging showed this
to be of fairly constant surface brightness, sharply bounded on the NE
side, and much more ``fluffy'' and rather ill-defined on the SW side
close to the main emission region. The tip of the filament is
cospatial with a small galaxy suggesting a causal connection. In
Figure \ref{fig8} we present the spectrum of this filament obtained
with the ESI instrument.

\begin{figure}[t]
\epsscale{1.0}
\plotone{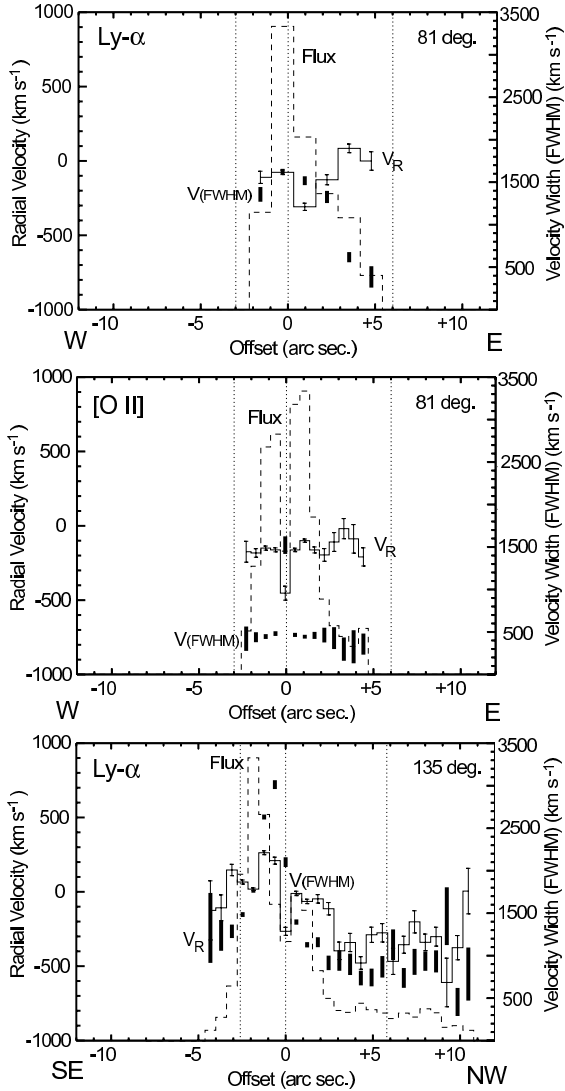}
\caption{\label{fig7} Similar to fig. \ref{fig3} but for \squ. \lya\
(top) and \oii\ (middle) emission along PA=81\degree\ and \lya\
emission along PA=135\degree (bottom).}
\end{figure}

Figures \ref{fig6} and \ref{fig7} show that the filament is offset
from the systemic velocity of the radio source by approximately
$-$200\,\kmps. The central wavelength fit by a single Gaussian yields
$\lambda \approx 5818.1 \pm 0.3$\,\AA\ or $-175$\,\kmps\ relative to
the systemic velocity at $z = 3.789$. The line has a FWHM of $20 \pm
1$\,\AA\ corresponding to a (deconvolved) velocity dispersion of
$\approx 300$\,\kmps.  The fact that there is no evidence for a strong
($>100$\,\kmps) velocity gradient across the filament seems to
indicate that it must be moving in the plane of sky.

Observations with mm-interferometers of the redshifted CO J=4$-$3
transition in \squ\ have found two kinematically and spatially
separate gaseous reservoirs \citep{Papadopoulos00}. Recently this has
been confirmed by VLA observations of the CO J=1$-$0 transition
\citep{Greve05}. An overlay of the \lya\ image with the
mm-observations, \emph{see} fig. 5 in \citet{Reuland03}, suggests that
the filament extends in a direction perpendicular to the major axis of
the molecular gas, reminiscent of the (polar ejection) morphology of a
galactic superwind. The CO J=4$-$3 observations showed a gas-rich
narrow component with a FWHM of $\approx 150$\,\kmps\ at a velocity of
$-224$\,\kmps\ to the galaxy. This same component has now been
independently dectected by its CO J=1$-$0 emission at a relative
velocity of $-220 \pm 40$\,\kmps\ with a FWHM of 165\,\kmps.

As is the case for \tur, the optical spectroscopic observations show
similar velocities for the molecular and emission line gas. This
corroborates the idea of a superwind, as it provides a dynamical
connection between the filament and the gas fueling the nuclear AGN.
If this is the case, the source of the ionization may be either shocks
or photoionization from the central nucleus.

The superwind interpretation of the Ly$-\alpha$ filament does however
present certain difficulties. In particular, it is hard to understand
how this feature, over 70 kpc in length, manages to remain so
well-collimated and kinematically quiescent over such a distance if it
were related to a
high- speed outflow (albeit in the plane of the
sky). However, the filter cut-off may play an important role here,
since the the observed velocity of the filament is very close to the
filter cut off. Thus, it may give the appearance of a sharp boundary,
but may not be one in actuality. Possibly a slit placed over ``empty
sky'' next to the filament would reveal other portions of the filament
at different velocities. As alternative possibility may be that it
represents an accretion filament or a tidally stripped gas stream
extending from the active galaxy. In this case, the both the
quiescence of its velocity field, and the similarity of the systemic
velocities of the filament and of the molecular reservoir would both
find a more natural explanation. Photoionization by the central AGN
would then be the most probable cause of the ionization of the
filament. On balance, this accretion or tidal stripping scenario seems
to provide the most convincing explanation for the filament.

\begin{figure}[t]
\epsscale{0.9}
\plotone{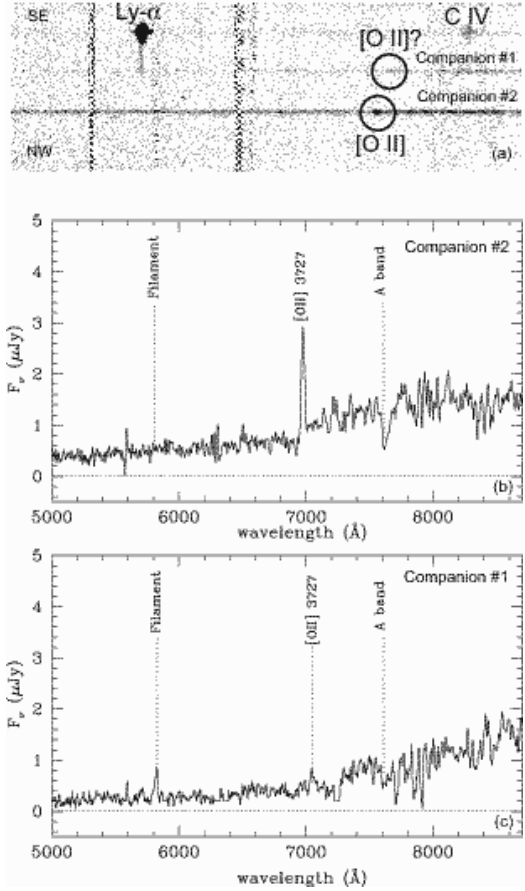}
\caption{\label{fig8} (a) A 2-D spectrum of the galaxy (\# 1) at the
tip of the \lya\ filament of \squ, together with the spectrum of a
brighter companion galaxy (\# 2) which happened to lie on the
slit. The position of the galaxy might suggest a physical association
with the \lya\ filament of \squ. However, the spectrum of companion
galaxy \#1 bears close resemblance to the brighter foreground galaxy
\#2 shown in panels (b) and (c), respectively.  The circles in panel
(a) indicate emission lines which are probably redshifted \oiil\ in
both foreground galaxies, identified in the spectra of panels (b) and
(c).}
\end{figure}

The galaxy at the tip of the filament is most likely to be a chance
superposition.  Figure \ref{fig8} shows the 2-D and 1-D spectra of
this galaxy (Companion \# 1), and a second companion galaxy (Companion
\#2) which also fell on the slit. The spectrum of Companion \#1 is
very similar to, but fainter than Companion \#2.  The brighter galaxy
shows an emission feature, identified as the [\ion{O}{2}] line and a
break near 7000\AA, while the overall shape of the spectrum for the
fainter galaxy is similar but without an evident break.  Identifying
the line at 7047.0\AA\ with redshifted \oiil\ yields a redshift of $z
= 0.891$ for the galaxy. This implies that it is a foreground object
and not causally connected to the filament, despite its suggestive
location.

\subsubsection{\hor: General Remarks}

\begin{figure}[t]
\epsscale{1.0}
\plotone{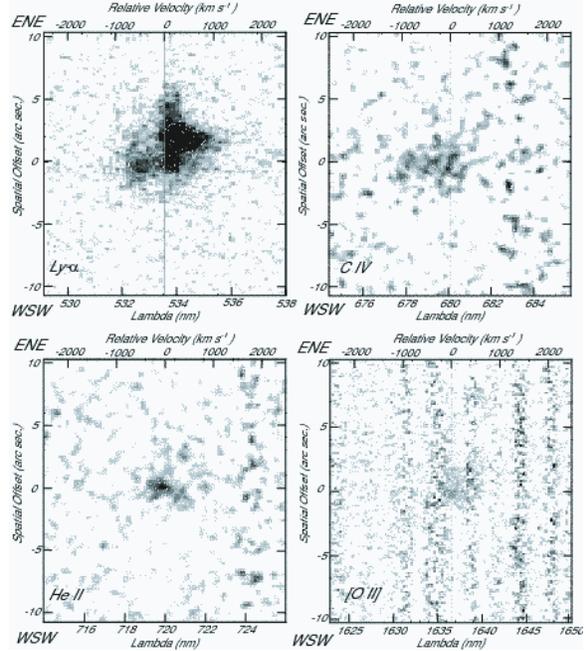}
\caption{\label{fig9} Similar to fig. \ref{fig2}, but for \hor. The
top left panel shows the line emission profile for \lya, top right
\civ, bottom left \heii, and bottom right \oii.}
\end{figure}

The optical spectrum obtained for \hor\ is more sensitive than
previous observations reported in the literature (Lilly 1988;
Martin-Mirones et al. 1995). We have detected \lya\ emission over an
extent of \mbox{$\approx 10\,\arcsec$} with a complex multi-component
spatial and velocity structure (cf. fig \ref{fig9}) and have detected
extended (3\arcsec ) continuum emission from the core over the entire
wavelength range with an almost constant surface brightness. The \lya\
profile shows galaxy-wide blue-shifted absorption by neutral hydrogen.

As is evident from Table \ref{linetable} and Figure \ref{fig9}, the
inferred redshifts depend on both the emission line and the position
of the aperture used. Much of this is due to the absorption optical
depth effects in the line. The \heii\ line at 7199.35 $\pm 0.5$\,\AA,
extracted in a 2\arcsec\ wide aperture centered at the radio core, is
probably most representative of the systemic velocity of the
galaxy. From this we infer a redshift $z = 3.3886 \pm 0.0003$. This is
slightly less than that reported previously by \citet{Lilly88} and
\citet{MartinMirones95}. They inferred $z = 3.395$ and $z=3.391$
respectively. However, these reported redshifts are entirely
consistent with redshift estimates we derive using other lines in the
same aperture, emphasizing the difficulty of assigning single
redshifts to such complex and self-absorbed systems.

\subsubsection{\hor: Kinematics}\label{hor:kinematics}

The \heii\ line is only a few hundred \kms\ wide and may be slightly
extended, with its position of maximum surface brightness centered at
the position of the radio nucleus. We can therefore argue that it
arises from photo-ionization of the dense gas surrounding the
nucleus. The \oii\ line is broad and spatially extended, but is too
faint to allow a detailed comparison with the kinematics of the other
lines.

Given that \hor\ shows associated \hone\ absorption against the radio
continuum it is important to see whether this is present also in the
\lya\ profile. This emission line shows a self-absorption feature on
the blue side of the profile centered at $\lambda = 5331.58$ ($z =
3.3857$), and which can be fitted with a Voigt profile with a column
density of $1.8 \times 10^{14}$\,cm$^{-2}$ and Doppler parameter $b =
195 \pm 11$ \kmps.  The inferred blue shift is $\approx100$~\kms. This
cannot be the same gas that causes \hone\ absorption against the radio
continuum, since \citet{Briggs93} and \citet{CodyBraun03} infer a
redshift $z_{\rm abs} = 3.3962$ with $\nh = 3 \times
10^{21}$\,cm$^{-2}$ and FWHM = 120\,\kmps\ for the absorbing gas. It
is probable that the dense gas giving the \hone\ absorption lies close
to the nucleus, and any absorption that this may have produced in the
\lya\ line is completely veiled by the frequency re-distribution
produced by multiple scattering events.

The \lya\ absorption can possibly be identified with a galaxy-wide
outflow of material. In this respect, it is very similar to the
galaxy-wide outflow identified by \citet{Wilman05} in a \lya\ galaxy
associated with the SSA-22 protocluster. We will return to this point
in the discussion.

\begin{figure}[t]
\plotone{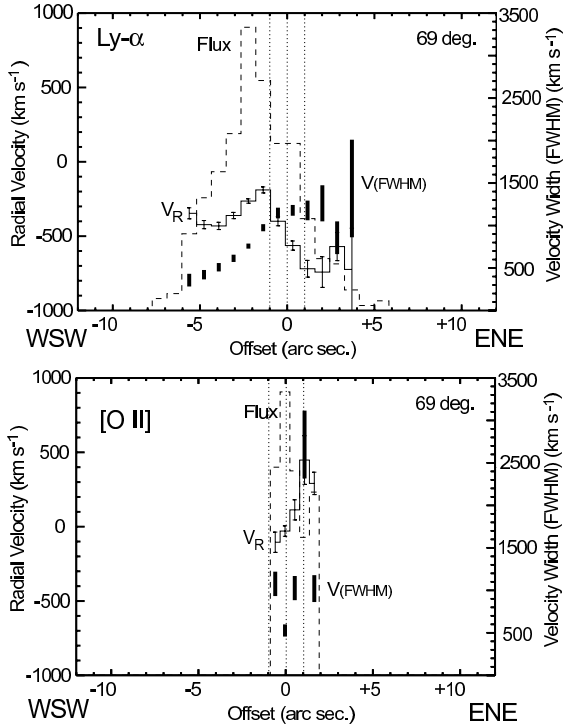}
\caption{\label{fig10} Similar to fig. \ref{fig3} but for \hor. \lya\
(top) and \oii\ (bottom) both along PA=69\degree.}
\end{figure}

The underlying \lya\ emission profile revealed by fig. \ref{fig9} is
very interesting, as it shows evidence of a strong velocity shear
amounting to nearly 1000 \kms\ across the central arc sec. The upper
(NE) component has the typical triangular shape resulting from
spatially extended absorption that is blue-shifted relative to the
systemic velocity \citep[cf.][]{Dey99}, and it is terminated by the
self-absorption on the blue side. The lower (SW) component may well
have an underlying shape which is the symmetrically reversed version
of the NE component, but the self-absorption makes this difficult to
trace, and the line reaches its maximum brightness at a relative
velocity of $-500$\,\kmps, as it emerges from under the
absorption. Because of the self absorption, a gaussian fit to the line
profile is not valid, and yields spuriously large values, as can be
seen in fig. \ref{fig10}.

The \civl\ emission is both spatially extended and broad ($\approx
1000\kms$) and is centered on the nucleus. The line broadening in the
nucleus is probably, once again, the result of a direct interaction
between the relativistic outflow associated with the peculiar radio
source \citep[see ][for details]{Carilli95} and its dense cocoon of
interstellar medium.

\section{Discussion \label{Discussion}}

\subsection{\label{Elines}Emission Line Diagnostics}

The relative intensities of emission lines are, in principle, powerful
diagnostic tools for studying the excitation mechanisms, metallicities
and physical conditions in the emission line gas associated with
AGN. Diagnostic diagrams for optical line ratios have been most
extensively used to probe the gas in nearby active galaxies, where
evidence for both jet- and accretion- powered shocks and for
photoionization by the central AGN has been adduced
\citep[\eg][]{Bicknell97,Dopita97,Groves04a,Groves04b}. Consequently
these relationships have been best calibrated for rest-frame optical
lines. However, line diagnostic diagrams applicable to both shock- and
photo-ionized AGN have been developed for use in the UV
\citep{Allen98,Groves04b}.

Until recently, for most \hzrgs\ \citep{DeBreuck00} or for the 3C
radio galaxies (Best, R\"ottgering, \& Longair 2000a,b; Inskip et
al. 2002a,b) only the rest-frame UV lines have been available. The
optical line diagnostics have previously not been commonly used
because of calibration difficulties and the limited sensitivity of
near-IR spectrographs on telescopes of the 2-4\,m class.

We would like to be able to use both UV and optical emission line
diagnostics to determine the mechanism that is responsible for
ionizing these extended emission line nebulae. The most likely
candidates are: photoionization by radiation from an AGN or stars,
shock heating, and shocks with precursors that photoionize the region
ahead of the shock by radiation from the gas. Studies of $z \approx 1$
radio galaxies have shown that the dominant ionization mechanism may
depend on the evolutionary state of the radio source \citep[\eg\
][]{Best00a,Best00b,Inskip02a,Inskip02b}. Generally, it was found that
the emission line gas of small (i.e. young) sources is shock ionized,
but as the radio source expands beyond the host galaxy, interactions
with the gas decrease and photoionization by the AGN takes over.  As
\citet{DeBreuck00} noted, for \hzrgs\ the situation may be more
complex.

Before we investigate these line diagnostics, we shall first use
global energy requirements to constrain important source parameters on
the basis of both photoionization and shock models.  The total \lya\
fluxes of the extended emission line regions are of order $10^{-14}
\ergpspcm$, yielding luminosities $L_{ \rm Ly\alpha} \approxeq 10^{45}
\ergs$ (\citet{Reuland03}). If the source of ionization is an active
nucleus, then the luminosity of the source derives either from the UV
photons which it produces, or from the mechanical energy of the jets
which it powers.  These luminosities can be derived from the
theoretical ratio of the Ly$-\alpha$ flux to the total flux in both
shock and photoionization models. To do this we use the shock models
from \citet{Dopita96}, and assume that the PdV work inmplied by this
luminosity is the mechanical energy flux of the jet. For the
justification of this assumption, see \citet{Bicknell97}. For
photoionization, we use the radiation-pressure dominated
photoionization models from Groves, Dopita, \& Sutherland (2004a,b),
assuming a power-law spectrum with index $\alpha = - 1.4$ ($f_{\nu}
\approx \nu^{\alpha}$). From these we derive either an ionizing
radiative or a mechanical energy flux of at least $1.1 \times
10^{46}$\ergs.  This flux is consistent with a typical luminous
embedded QSO.

Alternatively, some of the ionization may be due to young stars such
as are found in the numerous clumpy components of the Southern region
of \tur\ \citep{vanBreugel99}, and whose formation has probably been
triggered in the dense cocoon surrounding the central radio jet
\citep{Dey97,Bicknell00}. In this case, assuming a stellar radiation
field with an effective temperature of 42,000\,K, we infer (using the
\citet{Kennicutt98} calibration) that a star formation rate of at
least $3000$\,\Msunpyr\ would be needed to ionize the halo. This star
formation rate is remarkably close to what is inferred from rest-frame
far-IR observations \citep{Dunlop94,Stevens03}. However, this does not
necessarily imply that stellar photoionization is the dominant
ionization mechanism. Photoionization by stars usually results in
\lya\ equivalent widths less than 240\,\AA\ \citep[assuming stellar
populations at solar metallicity;][]{CharlotFall93} and would seem
inconsistent with the large equivalent widths observered
\citep[$\approx 500$\,\AA;][]{Dey97}. The strength of the \civ\ lines
and other high-excitation species also militates against a purely
stellar origin of the ionizing flux. Indeed, \cite{Bicknell00} have
used the emission spectrum to infer that shocks are the dominant
source of ionization in this region, and have used the flux to
determine the $PdV$ work being done by the relativistic jet. However,
away from the radio axis photoionization by stars may well become more
important.  \tur\ is a young system and recent modeling suggests that
any primordial stellar populations could be very efficient in emitting
significant amounts of ionizing flux while remaining virtually
undetectable in the optical \citep[\cf][]{Fosbury03,Panagia03}.

Using the observed line ratios, better constraints on the ionizing
process can be derived. \citet{Dey97} determined the following ratios
for the narrow rest-frame UV lines of \tur: \ciii/\civ\ $\approx$ 0.14
($\approx$ 0.7 for total fluxes) and \civ/\heii\ $\approx$ 2.4. The
latter is close to the maximum of 3.1 predicted by photoionization
models.  Although the narrow emission line strengths of \ciii, \civ,
and \heii\ can be matched by a simple nuclear AGN photoionization
model with solar metallicity clouds, a high ionization parameter $U
\la 0.1$, and a power-law ionizing source with index $\alpha = - 1.5$,
\citet{Dey97} and \citet{Bicknell00} argue that the dominating
ionizing mechanism is shocks rather than photoionization the nucleus.

\begin{figure}[t]
\plotone{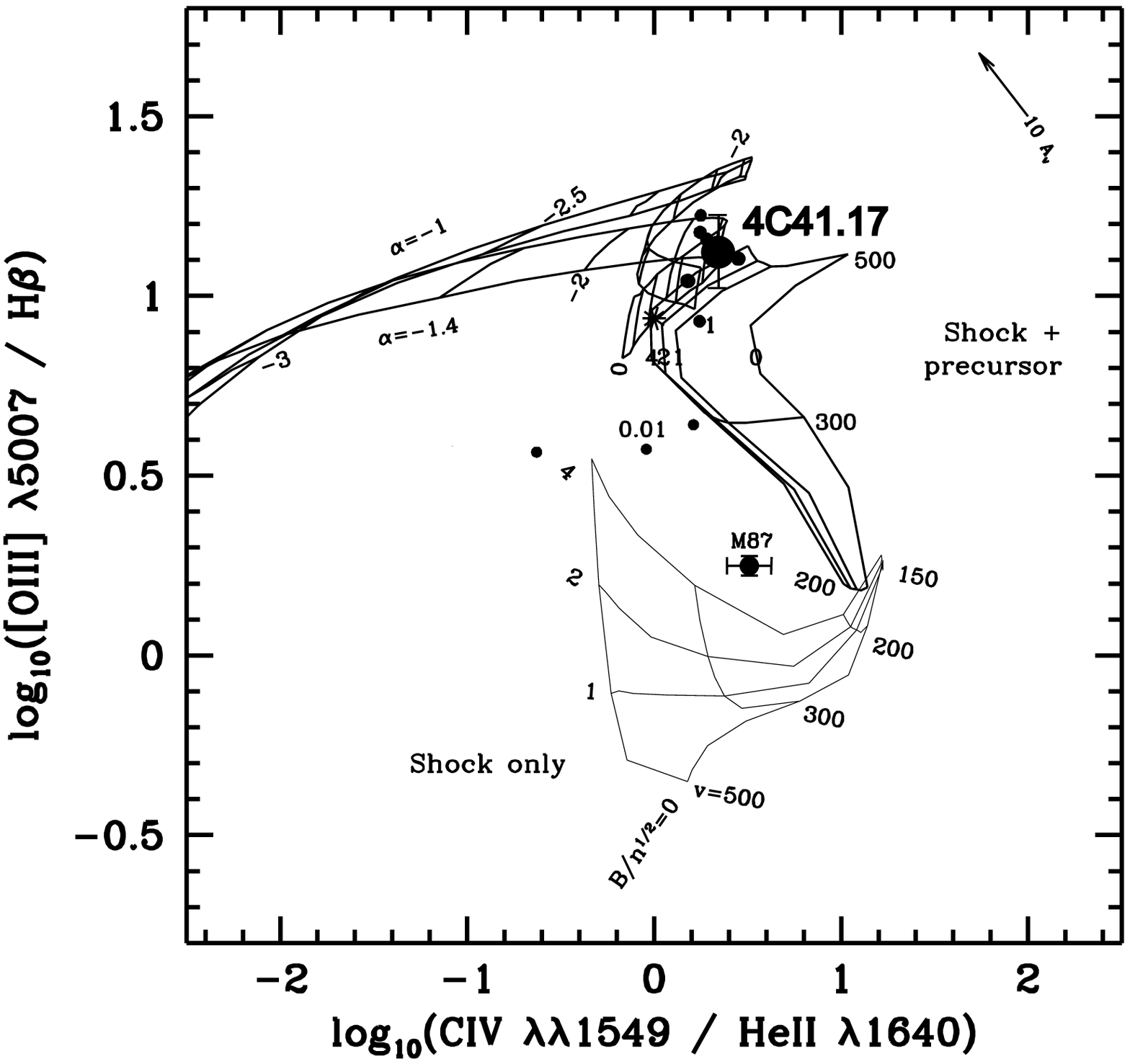}
\caption{\label{fig11}A UV-optical line diagnostic diagram adapted
from \citet{Allen98}. The ratios \oiii/\hbeta\ and \civ/\heii\ are
used to separate shock (lower grid), shock+precursor(right hand side)
and photoionization (top left) models for solar metallicity gas. The
shock models are marked according to the shock velocity and magnetic
parameter, and the photoionization models according to the ionization
parameter, {\cal U}, and the power-law index of the photoionizing
radiation field ($\alpha$). The points represent nearby $z\approx1$
Galaxies, and the accretion disk of M87 in explicitly identified.  The
large point with error bars is for the nucleus of \tur. This proves
that the excitation of \tur\ is similar to the other radio galaxies,
and like them is shock-dominated.\label{ionizationdiagram}} \
\end{figure}

Figure \ref{fig11} shows that a combination of rest-frame UV
(\civl/\heiil) and rest-frame optical (\oiiil/\hbeta) emission line
flux ratios can help to separate pure photoionization from shock
dominated mechanisms. This diagram was derived assuming solar
metallicity gas and is relatively insensitive to the effects of dust
extinction as the ratios are determined from lines close in
wavelength.  \citet{Iwamuro03} conducted a rest-frame UV-optical
emission line study of 15 radio galaxies with $2 < z < 2.6$.  They
found that there is a range in observed line-ratios suggesting that
some objects are best explained with photoionization of low
metallicity gas while others are consistent with the shock+precursor
model. \citet{Carson01} and \citet{Maxfield02} found evidence for
changing line ratios {\it within} sources, suggesting that the
dominant ionization process are depends on the region of interest.

Based on previously published values of \oiii / \hbeta\ $\approx$ 3.4
and 2.8 for \tur\ and \hor\ respectively \citep{EalesRawlings93} it
seemed that pure photoionization models could be ruled out. However
the \hbeta\ detections were marginal, and using the total integrated
line fluxes for \tur\ from the present study we derive ratios of
\oiiil/\hbeta\ $\approx$9.6 for the central region and \oiiil/\hbeta\
$\approx11.8$ for the spatially integrated spectrum that are much
higher and are consistent with both pure photoionization and
shock+precursor models. The point derived for the spatially integrated
spectrum of \tur\ is also shown on fig. \ref{fig11}. This appears to
show that the excitation of \tur\ is similar to other radio
galaxies. It cannot be due predominantly to stars, for which the
\oiii/\hbeta\ would be of order three and for which both \civl\ and
\heii\ would be very faint.

\begin{figure}[t]
\plotone{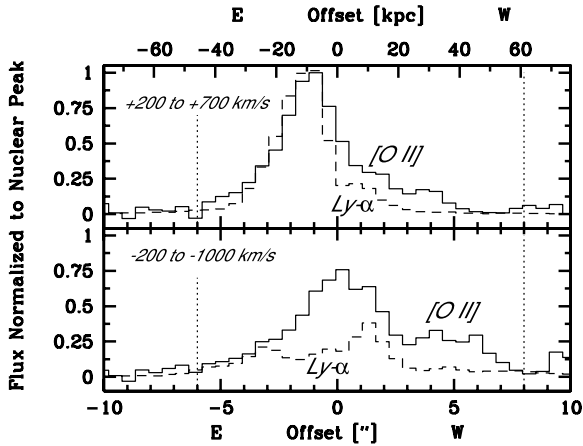}
\caption{\label{fig12}Top: Normalized surface brightness profiles of
the \oii\ (solid line) and \lya\ (dashed line) emission along the
inner radio axis and south-west filament of \tur. These are integrated
in the velocity range $+200$ to $+700$\kms and clearly show the
spatially-extended \oii\ emission.  The spatial zero-point corresponds
to the position of the radio core.  Bottom: Similar to top panel but
for the velocity range $-200$ to $-1000$\kms.  The \oii\ emission is
detected out to $\approx 8$\arcsec\ ($\approx 60$\kpc) west of the
nucleus, where it is blue-shifted by $\approx -600$\kmps\ relative to
the \heii\ line.  The projected distance of the south-west and
north-east radio lobes along the slit direction are indicated (dotted
lines).  The $+200$ to $-200$\kms\ range for \oii\ is confused by
near-infrared sky lines and is therefore not shown.}
\end{figure}

The extended regions of \tur\ have a much lower excitation,
however. As discussed in \S \ref{4c41results}, for \tur, \oii\
emission was detected as far as $\approx 60$\kpc\ from the
nucleus. Figure \ref{fig12} shows the velocity integrated relative
intensities of \lya\ and \oii\ as a function of distance from the
nucleus for a red-shifted ($+200$ to $+700$\,\kmps) and a blue-shifted
($-200$ to $-1000$\,\kmps) velocity interval.  The red-shifted \oii\
flux closely follows that of the \lya\ , while the blue-shifted \oii\
shows a relative enhancement over the range 3--8\arcsec\ W of the
nucleus. In this region no evidence for \oiii\ is found
(Fig. \ref{fig2}), whereas the \oiii\ emission in the nuclear region
is much brighter than the \oii. This shows that \oiii/\oii\ ratio is
much lower ($\approx$1--2) in the outer regions than in the center
($\approx$3--4) implying that a different mechanism is responsible for
the extended \oii\ emission. The \oiii/\oii\ ratio is known to be
sensitive to the ionization parameter in the nebula. Because they have
higher ionization parameters and also have central stars with higher
effective temperatures for a given age, the low-metallicity \HII\
regions are characterized by higher \oiii/\oii\ ratios. Since this
ratio falls to very low values in the outer parts of \tur\, this
proves that Population III stars with their very high effective
temperatures cannot be responsible for the excitation of the outer
regions. Excitation by shocks is a more likely ionizing source, since
the emission is extended along the radio-axis and both the \lya\ and
\oii\ are blue-shifted by $\approx600$\,\kmps.  Furthermore, the peak
in the \oii\ emission on the W side of \tur\ corresponds to the point
at which the velocity dispersion drops steeply and boundary of the
outer radio lobe. This suggests that the \oii\ is excited by shocks in
an outward-propagating cocoon around the radio lobe. We note here that
\oii\ emission is particularly strong in shocks. This corroborates the
suggestions by \citet{Dey97} and \citet{Bicknell00} that the \lya\
emission, at least in this direction, is not due to scattering but
that the nebula must be locally ionized by shocks related to radio
source. \citet{Scharf03} proposed that the escape of X-ray emission
from the shocked region could be responsible for ionizing the extended
halo beyond the the radio lobes.

From the jet shock spectrum, \citet{Bicknell00} inferred that the
metallicity of the star-forming cocoon around the radio jet was about
the same as for the LMC; a half to a third of solar. For the nucleus
itself, \nv\ is often used as a tracer of the chemical abundance,
because it is rather insensitive to shock ionization, density
fluctuations, ionization parameter, and as a secondary nucleosynthesis
element has a quadratic dependence on the metallicity
\citep{VillarMartin99,Groves04a}.  This would imply that that would
only be detected in highly chemically enriched sources, and not
significant in the majority of \hzrgs\
\citep{Rottgering97,DeBreuck00}. \citet{Vernet01a} presented a
metallicity sequence for \hzrgs\ that is similar to the one for
quasars \citep{HamannFerland93}. \tur\ was included in their study and
a metallicity of $Z$ $\approx$ 1.3 \Zsun\ was inferred for the central
region \citep[aperture of $2\arcsec \times 1\arcsec$][]{Dey97}.

Thus, the nuclear and jet cocoon spectra of \tur\ argue for a fair
degree of chemical enrichment, perhaps more in the nucleus than in the
jet region. However, the halo region is by no means primordial in its
abundances. Although we cannot directly infer the actual metallicity,
we can say that the detection of strong \oii\ emission out to $\approx
60$\kpc\ from the nucleus demonstrates that the halo has been
chemically enriched throughout its volume, at least in the illuminated
regions.

A similar conclusion is also inferred for \squ. The very similar
velocity morphologies and spatial extent of the \lya\ and \oii\
spectra at PA = 81\degree\ seen in figure \ref{fig5} once again shows
that the enriched gas is distributed throughout the \lya\ halo. The
strongly- \civ\ emitting shell in the nuclear vicinity shows that this
galaxy too is excited primarily by the AGN rather than by star
formation. The expanding shell morphology of both \lya\ and \civ\
further argues that \squ, like \tur, is excited mostly by shocks in
the dense expanding cocoon of interstellar gas pushed by the
relativistic jets from the nucleus.

\subsection{Galaxy Mass estimates}

The \hzrgs\ are known to represent some of the most massive galaxies
galaxies formed in the early universe. An estimate of their mass, by
any means, is important to constrain the $\Lambda$-CDM models of
galaxy formation, since the formation of such massive galaxies so
early on might present something of an issue for the theory to
resolve. We therefore provide these mass estimates in this section.

\subsubsection{Luminosity based masses}

\citet{vanBreugel99} estimated the mass of \tur\ using the integrated
rest-frame UV light. They found it to be a very massive galaxy with
$L_{\rm UV} \approx 17 L^{*}$.  Similarly, \citet{Graham94} used
near-IR imaging with the line-free $K_{S}$ filter to estimate the mass
of the galaxy directly. In a 4\,\arcsec\ diameter aperture they found
$K_{S} = 19.6 \pm 0.6$\,mag. This converts in their cosmology to an
absolute magnitude of \tur\ to be $M_{\rm B} = -23.0$, or about 20
$L^{*}$.

A lower mass limit for the ionized gas in the halo can be obtained
from photoionization modelling of the total \lya\ flux, neglecting the
effect of dust absorption, and assuming that the ISM is fully
ionized. From this we derive an average thermal electron density in
the halo of $n_{e, halo} = 0.1\, f_{\rm v}^{-\frac{1}{2}}\,\rm
cm^{-3}$ where $f_{\rm v}$ is the volume filling factor of the \lya\
emitting gas. This yields a mass of gas of $M_{\rm halo} \approx 6
\times 10^{12} \, f_{\rm v}^{\frac{1}{2}} ~\rm M_\odot$. For a volume
filling factor $f_{\rm v} \approx 1$ the inferred amount of gas is
substantial and comparable to the broad-band mass estimates of the
host galaxy and that of fully formed, massive elliptical galaxies in
the local Universe. However, the fact that the \lya\ line appears to
provide a good dynamical mass tracer, and is not grossly affected by
self-absorption argues in favor of $f_{\rm v}<<1$, and consequently,
for a much lower total mass. The molecular component of \tur\ has
recently been traced by \citet{DeBreuck04}, who have detected the
CO(4-3) transition using the IRAM interferometer. They find two
massive CO clouds ($M_{\rm dyn}^{\rm rot} \approx 6 \times
10^{10}$\,\Msun) which coincide with two different dark lanes located
close to the nucleus in the deep \lya\ image. This is similar to the
CO content of the sub-millimetre galaxies, as measured by
\citet{Greve05} ($M_{\rm CO} \approx 3 \times 10^{10}$\,\Msun),
suggesting once again that the \hzrgs\ and the sub-millimeter galaxies
are drawn from similar populations of galaxies, differing only in the
activity of their AGN.

\subsubsection{Dynamical mass estimates}

\citet{vanOjik97} and \citet{VillarMartin03} have presented mass
estimates for the more quiescent parts of the emission line regions in
their studies.  There are two methods of estimating the dynamical
masses. First, one can assume that the halos consist of gas that has
settled in rotating disks. Secondly, the halos can be envisaged as
consisting of virialized clumps, which have velocity dispersions that
balance the gravitational forces.

In the case of rotating disks, the mass can be estimated by measuring
the velocity shear across the halo and using: $M_{\rm dyn}^{\rm rot}=
{{R V^{2} } / {G \sin^{2} i}}$ with $R$ the radius of the disk, $V$
half the amplitude of the rotation curve, and $i$ the inclination of
the disk with respect to the plane of sky. \citet{VillarMartin03}
found evidence for rotation in about $\approx 50\%$ of the objects
they looked at and inferred masses of order $M_{\rm dyn}^{\rm rot}
\times \sin^{2} i \approx 0.3-3 \times 10^{12}$\,\Msun.

Looking at our measured relative velocity distributions for \tur, we
infer a velocity shear of $\approx 300$\,\kmps\ (+100\,\kmps to
$-$200\,\kmps) at PA = 76\,\degree, over a distance of R =
20\,\arcsec/2 = 77\,kpc. From this we infer a dynamical mass of about
$M_{\rm dyn}^{\rm rot} \times \sin^{2} i \approx 4 \times
10^{11}$\,\Msun. The velocity profiles at position angles PA
=19\,\degree, and 21\,\degree\ show no signs of a shear, consistent
with their approximate alignment along the axis of rotation.

The observations for \squ\ does not show an indication of being
settled in a disk. Rather, it could be supported against gravitation
by velocity dispersions of the cloudlets in the halo. For this
scenario the dynamical mass is given by: \hbox{$M_{\rm dyn}^{\rm vir}
= {{5 R V_{R}^{2}} / {G}}$}, with $V_{R}$ the radial velocity
dispersion of the clouds. Using this relation we derive dynamical
masses for the outer regions of $M_{\rm dyn}^{\rm vir} = 5 - 10 \times
10^{11}$\,\Msun.  For the inner regions, where the signal-to-noise
ratio is better we find lower mass of $M_{\rm dyn}^{\rm vir} \approx 2
\times 10^{11}$\,\Msun.

The \lya\ emission profile of \hor\ revealed by fig. \ref{fig9}
suggests a strong nuclear velocity shear. If this is interpreted as a
rotation, then it would imply a rotational velocity of $\approx
550$\,\kmps\ over a distance of 3\,\arcsec\ or a rotation velocity of
$\approx 275$\,\kmps\ at $R \approx 12$~kpc. This implies a dynamical
mass within this core region of $M_{\rm dyn}^{\rm rot} \times \sin^{2}
i \approx 2 \times 10^{11}$\,\Msun.  However, as discussed in section
\ref{hor:kinematics} this velocity gradient could equally well be due
to the expansion of a shell of interstellar gas driven by the a
directed bipolar outflow from the AGN with a velocity $v_{\rm exp}$.

\section{Outflows}

Absorbing gas associated with lower velocity outflows has now been
observed in several \hzrgs, almost all of which show asymmetric \lya\
profiles suggesting the presence of blue-shifted absorbing gas which
appears to be spatially extended over the entire emission line region
\citep[see \eg\ ][]{vanOjik97,Dey99,Jarvis03,Wilman04}.  Spectroscopic
evidence for outflows at high redshifts exists also for other,
presumably less massive, galaxies \citep{Pettini01, Dawson02}.

There is an ongoing debate whether massive ($> L_{*}$) galaxies or
less massive ($< L_{*}$) galaxies enrich the ICM.  \citet{Martin99}
and \citet{Heckman00} found that outflow speeds, $v_{\rm exp}$, are
largely independent of galaxy mass. This would imply that smaller
galaxies are more efficient at ejecting enriched material to large
radii. Specifically, \citet{Heckman02} claim that for massive galaxies
the metals will not escape the deep potential wells. However, recent
modelling suggests that the bulk of the metals in clusters are
produced by $L_*$ and brighter galaxies \citep[\eg\ ][]{Nagashima04}.

For all the three \hzrgs\ we have observed, we found features that can
be interpreted as symptomatic of outflows.  The extended optical
filament of \squ\ is consistent with an outflow in the plane of
sky. \squ\ also shows crescent shaped arcs around the radio core and
unobscured optical nucleus that are consistent with an expanding
cocoon; see fig. \ref{fig6}.  The blue-shifted carbon-rich component
of \hor\ seems to point towards an outflow of enriched material with
an outflow velocity, $v_{\rm exp}$ of up to 1000\,\kmps.  The
near-infrared [\ion{O}{2}] spectroscopy (Figs. \ref{fig2}, \ref{fig3},
and \ref{fig12}) and the earlier optical \lya\ spectroscopy of \tur\
\citep[Fig. \ref{fig2};][]{Dey97}, show that both emission lines
exhibit large blue-shifted radial velocities $\approx 600 - 900\kms\ $
seen in projection along the radio axis.  In particular, the gas is
very disturbed along the south-west filament, with \lya\ velocity
widths ranging up to $\Delta v_{F} \approx 900 - 1600 \kms\ $.  Beyond
the radio hotspot, the velocity and velocity widths decrease abruptly.
The kinematics, its radial filamentary structure, and chemical
enrichment of the gas all indicate a process of entrainment of
material away from the central regions by the radio jet.  In this
scenario, the optical filament represents the shocked radiative cocoon
of the radio source.

For a galaxy mass of $M_{\rm gal} \sim 1\times 10^{12}$\,\Msun within
a radius of $R \sim 40$kpc, figures that seem to be typical for these
\hzrgs, the escape velocity $v_{\rm esc}=(2GM/R_{\rm gal})^{1/2}$ is
of order $\sim 450\,\kmps$. The velocity gradients found from the
spectroscopic observations range a factor 1.5--2.5 larger than this,
and implies that the ejection of the ISM originating from close to the
nucleus of such galaxies is a viable scheme to both chemically enrich
the inter-galactic medium and to terminate the epoch of star formation
in these massive galaxies.

While it may still be true that purely starburst powered super
galactic winds may not be sufficiently energetic for enriched nuclear
material to escape from the galaxian potentials, the additional
driving force produced by the energy content of the relativistic lobes
produced by the central AGN seem quite able to blow matter out of the
potential of the galaxy. This is direct evidence of the operation of
the type of feedback process proposed by \citet{SilkRees98}. In this,
the growth of the super-massive central black hole is terminated by
the ejection of the reservoir of interstellar gas, once the mass of
the black hole exceeds a critical value. In their model, this critical
value is determined by the relationship between the radiation pressure
and the depth of the stellar potential, and it is able to provide a
qualitative description of the Magorrian relationship
\citep{Magorrian98}, or the relationship between the mass of the
central black hole and the velocity dispersion of the bulge
\citep{Magorrian98,FerrareseMerritt00,Gebhardt00}.

The observations presented here suggest that this model needs to be
modified somewhat. Rather than being driven by radiation pressure, the
dynamical evidence is that they are driven by the $PdV$ work done on
the galactic medium by the relativistic jets, and that the ejection is
occurring in the faster moving material in the shocked cocoon around
the radio jet. Such a scheme is also suggested by the hydrodynamical
simulations of \citet{Springel05}. However, rather than being ejected
isotropically in a disk-like galaxy, as in their model, in the \hzrgs\
directional jets are being driven into a more spheroidal distribution
of galactic interstellar gas. Given that \tur\ has clear evidence of
large amounts of shock-triggered star formation in this shocked
cocoon, we can conclude that, in the \hzrgs\ we are witnessing the
fireworks that terminate the epoch of star formation in these massive
galaxies -- the last hurrah before these galaxies become massive ``red
and dead" Elliptical galaxies that we find today in the cores of the
most massive clusters.

The multiple component, asymmetric, and twisted radio structure of
\tur\ suggests that during this phase, the central SMBH has
experienced multiple periods of radio source activity and precession,
\citep{vanBreugel99,Steinbring02}. The precessing radio source will
gradually evacuate the central region of the galaxy of its
interstellar medium.  The time scales for star formation and radio
source activity in \tur\ are very similar
\citep{Chambers90,Bicknell00} and comparable to that for transporting
the \oii\ gas along the filament out to the vicinity of the south-west
hotspot ($\approx 7 \times 10^7$ yrs at $\approx 900 \kms$). It
suggests an overall picture where the initial growth of the galaxy
through merging, the feeding and growth of the central black hole, the
growth of the stellar component, and finally triggering of SMBH the
outflow, and its associated starburst activity are all closely coupled
processes.

\section{Conclusions}

Extended \lya\ halos are observed both in \hzrgs\ and in the
sub-millimetre galaxies. Indeed, these objects may be drawn from the
same underlying population, with the distinction that the \hzrgs\
represent the radio-loud and active phase of the galaxian evolution.
This seems supported by the observation that the largest radio-quiet
\lya\ halos have similar sizes and only slightly lower luminosities as
the \hzrgs\, but lack their large multi-component continuum
structures.

Our long slit optical spectra of \tur\ , \squ\ and \hor\ have shown
that their \lya\ halos exhibit disturbed kinematics, with broad lines,
large velocity shears, and, in some cases, expanding shells associated
with the radio lobes. This clearly demonstrates that the relativistic
jets are driving strong shocks into the galactic medium.

In \tur\ the near-IR spectra reveal very extended \oii\ and \oiii\
emission distributed along the radio source axis, as far as 60 kpc
from the nucleus. This provides direct proof that the \lya\ halos are
both chemically enriched by star formation and ionized throughout the
majority of their volume. The hypothesis that the \lya\ is due to
scattering off \hone\ clouds is disproved. Likewise, the idea that the
halo gas is primarily composed of chemically pristine ``primordial"
gas is also disproved. However, we cannot be certain that the gas in
directions perpendicular to the radio axis contains a large primordial
component, because the sensitivity of our spectroscopic observations
is insufficient to detect Oxygen lines at the low surface brightness
expected. To settle this point would require very deep near-IR
spectroscopy of the halo gas {\it outside} the radio sources, well
beyond the hotspots and orthogonal to the radio axes. This would be a
very difficult observation to make.

Our observations of \hzrgs\ also help cast observational light on the
origin of the correlation found in galaxies between the stellar
velocity dispersion and the black hole mass
\citep{Gebhardt00,FerrareseMerritt00}. We find clear evidence of
outflows of chemically enriched gas associated with the jets, and
velocity dispersions in the expanding cocoons around the jets which
probably exceed the escape velocity. In addition, \hor\ shows \lya\
absorption can be identified with a galaxy-wide outflow of material,
similar to the galaxy-wide outflow identified by
\citet{Wilman05}. This demonstrates that tin the \hzrgs\ the central
black hole has grown sufficiently to be able to profoundly modify its
surrounding galactic medium, pushing it into outflow, and (in the case
of \tur ) triggering large quantities of star formation in the shocked
galactic medium cocooning the radio lobes. Furthermore, the precession
of the radio lobes seen in \tur\ will ensure that eventually, all of
the interstellar medium will be shocked and either be converted into
stars, or else ejected from the galaxy.

These observations can be understood in terms of the general picture
of \citet{SilkRees98} that feedback of the black hole on its host
galaxy eventually limits the growth of the black hole. However unlike
the \citet{SilkRees98} concept, the feedback process is not primarily
radiation pressure but is the mechanical energy input delivered by the
relativistic jets.  Thus, in the \hzrgs\ we may be observing the
moment where galaxy collapse gives way to mass ejection, a newly born
galaxy is revealed. We may speculate that this is the defining moment
where the such galaxies enjoy one last violent burst of shock-induced
star formation before beginning their evolution to become the ``red
and dead" massive Elliptical galaxies we see in our local universe.

\begin{acknowledgements}

We thank all staff at the W.M. Keck Observatory for their excellent
support. The authors wish to recognize and acknowledge the very
significant cultural role and reverence that the summit of Mauna Kea
has always had within the indigenous Hawaiian community. We are most
grateful to have the opportunity to conduct observations from this
mountian.  M.R. thanks Mario Livio and the Space Telescope Science
Institute for generous hospitality. This work was performed under the
auspices of the U.S. Department of Energy, National Nuclear Security
Administration by the University of California, Lawrence Livermore
National Laboratory under contract No. W-7405-Eng-48. WvB acknowledges
support for radio galaxy studies at the Institute for Geophyhsics and
Planetary Physics at Lawrence Livermore National Laboratory and at UC
Merced, including the work reported here, with the Hubble, Spitzer and
Chandra space telescopes via NASA grants HST GO-9779, GO-10127, SST
GO-3482, SST GO-3329 and Chandra/CXO GO-06701011.  M.D. acknowledges
the support of the ANU and the Australian Research Council (ARC) for
his ARC Australian Federation Fellowship, and also under the ARC
Discovery projects DP0208445 and DP0664434. This work was supported by
the European Community Research and Training Network ``The Physics of
the Intergalactic Medium''. AD's research is supported by NOAO, which
is operated by the Association of Universities for Research in
Astronomy (AURA), Inc. under a cooperative agreement with the National
Science Foundation.

\end{acknowledgements}

\clearpage

\renewcommand{\baselinestretch}{1.0}
\renewcommand{\arraystretch}{.6}

\begin{deluxetable}{cccc}
\tablenum{1}
\tablewidth{13cm}
\tablecaption{\label{Table1}Radio galaxy sample}
\tablehead{
\colhead{Source} & \colhead{RA (J2000)} & \colhead{DEC (J2000)} & \colhead{$z$}} 
\startdata
\hor & 09 05 30.11 & 34 07 55.9 & 3.389 \\
\squ & 05 12 55.17 & 60 30 51.1 & 3.789 \\
\tur & 06 50 52.14 & 41 30 30.7 & 3.798 \\
\enddata
\tablecomments{Positions of the radio core
\citep{Carilli94,Carilli97,Carilli95} and the systemic redshifts based
on the \heii\ line are adopted as the frame of reference in this
paper.}
\end{deluxetable}

\setlength{\hoffset}{-15mm}
\begin{deluxetable}{lcllcccccr}
\tablewidth{190mm}
\tablenum{2}
\tabletypesize{\footnotesize}
\tablecaption{Summary of observations and instrumental setups sorted by position angle (P.A.)\label{Table2}}
\tablehead{\colhead{Object} & \colhead{Date} & \colhead{Instrument} & \colhead{Setup} & 
           \colhead{Seeing} & \colhead{Slit} & \colhead{Resln.} & 
           \colhead{$\lambda$ Coverage} & \colhead{P.A}. & \colhead{Obs.Time}\\
           \colhead{} & \colhead{} & \colhead{} & \colhead{} & 
           \colhead{(\arcsec)} & \colhead{(\arcsec)} & \colhead{(\AA)} & 
           \colhead{(\AA)} & \colhead{(deg)}. & \colhead{(s)} 
 }
\startdata
\hor & 16/01/02  & LRIS    & LS 600/5000  &  0.9 & 1.5  &\phantom{0}4 &  5150$-$7650       &  68.7             & 2$\times$1800     \\
     & 07/01/02  & NIRSPEC & Low Disp. N5 &  0.9 & 0.76 & 14 & 14600$-$17400      &  68.5             & 6$\times$900      \\
\squ & 15/01/02  & LRIS    & LS 600/7500  &  0.7 & 1.5  &\phantom{0}4 & 5450$-$7600      &  81.0             & 2$\times$1800     \\
     & 25/02/01  & ESI     & Low Disp.    &  0.8 & 1.0  &          13\tablenotemark{a} &  4000$-$9600 &  135.2\phantom{0} & 3.5$\times$1800   \\
     & 07/01/02  & NIRSPEC & Low Disp. N6 &  0.9 & 0.76 &          14 & 15600$-$19800      &  81.4             & 6$\times$900      \\
\tur & 24/02/01  & LRIS    & LS 600/7500  &  0.7 & 1.5  &\phantom{0}6 & 5200$-$6100      &  19.4             & 3$\times$1800     \\
     & 23/02/01  & LRIS    & MOS 300/5000 &  0.6 & 1.0  &\phantom{0}4 & 4300$-$6600       &  21.6             & 4$\times$1800     \\
     & 10/12/96  & LRISp    & LS 400/8500  &  0.9 & 1.0  &\phantom{0}8 &  5500$-$9280       &  76.5             & 28$\times$1200\tablenotemark{b}\\
     & 15/01/02  & LRIS    & MOS 400/8500 &  0.7 & 1.0  &\phantom{0}6 & 5300$-$8000      &  81.0             & 3$\times$1800     \\
     & 03/02/97  & LRIS    & LS 600/5000  &  0.9 & 1.0  &\phantom{0}5 &  4320$-$6850       &  170.8\phantom{0} & 3$\times$1800     \\
     & 07/01/02  & NIRSPEC & Low Disp. N6 &  0.9 & 0.76 &          14 & 15600$-$19800      &  42.7             & 3$\times$900      \\
     & 07/01/02  & NIRSPEC & Low Disp. N6 &  0.9 & 0.76 &          14 & 15600$-$19800      &  70.2             & 5$\times$900      \\
     & 07/01/02  & NIRSPEC & Low Disp. N7 &  0.9 & 0.76 &          14 & 20300$-$25000      &  42.7             & 2$\times$900      \\
     & 07/01/02  & NIRSPEC & Low Disp. N7 &  0.9 & 0.76 &          14 & 20300$-$25000      &  67.8             & 4$\times$900      \\
\enddata
\tablenotetext{a}{Near the redshifted \lya\ line at $\lambda =
5825$\,\AA; the resolution varies roughly linearly from 3\,\AA\ at
3900\,\AA\ to 40\,\AA\ at 11000\,\AA.}
\tablenotetext{b}{Spectropolarimetric observations, a detailed
analysis of which was presented in \citet{Dey97}.}
\end{deluxetable}

\clearpage

\begin{deluxetable}{llccccccccl}
\tablenum{3}
\rotate
\tablewidth{220mm}
\tabletypesize{\tiny}
\tablecaption{Summary of Emission-Line Measurements\label{linetable}}
\tablehead{\colhead{Object} & \colhead{Line} & \colhead{$\lambda_{\rm
rest}$} & \colhead{$\lambda_{\rm obs}$}& \colhead{$z$} &
\colhead{FWHM} & \colhead{FWHM} & \colhead{Continuum Flux} &
\colhead{Line Flux} & \colhead{EW$_{\rm obs}$}\\ \ & \colhead{} &
\colhead{(\AA)} & \colhead{(\AA)}& & \colhead{(\AA)}&
\colhead{(\kmps)} & \multicolumn{2}{c}{($10^{-17}$\ergpspcm)} &
\colhead{(\AA)}}
\startdata
\hor\ (core)  
      & \lya         & 1215.7 & 5334.69 $\pm$ 0.09 & 3.3883 $\pm$ 0.0001 & 21.9 $\pm$ 0.2 & 1233 $\pm$  13 & 18.3 $\pm$  0.5 &  7890 $\pm$  95 &  431 $\pm$  12 \\
      & C IV(doublet)& 1549.0 & 6795.00 $\pm$ 0.62 & 3.3867 $\pm$ 0.0004 & 22.7 $\pm$ 1.5 & 1003 $\pm$  65 & 16.3 $\pm$  0.4 &  1770 $\pm$  70 &  109 $\pm$   5 \\
      & He II        & 1640.5 & 7199.37 $\pm$ 0.37 & 3.3886 $\pm$ 0.0002 & 17.2 $\pm$ 0.9 &  718 $\pm$  37 & 11.8 $\pm$  0.4 &  1502 $\pm$  60 &  127 $\pm$   7 \\

\hor\ (total) 
      & \lya         & 1215.7 & 5336.26 $\pm$ 0.23 & 3.3896 $\pm$ 0.0002 & 21.4 $\pm$ 0.4 & 1203 $\pm$  24 & 27.4 $\pm$  3.8 & 25640 $\pm$ 600 &  937 $\pm$  133 \\
      & C IV(doublet)& 1549.0 & 6799.58 $\pm$ 0.72 & 3.3897 $\pm$ 0.0005 & 10.4 $\pm$ 1.7 &  460 $\pm$  75 & 21.6 $\pm$  2.9 &  2590 $\pm$ 420 &  120 $\pm$   25 \\
      & He II        & 1640.5 &     &   \\
              
\squ\ (core)  
      & \lya         & 1215.7 & 5826.52 $\pm$ 0.07 & 3.7928 $\pm$ 0.0001 & 47.1 $\pm$ 0.2 & 2424 $\pm$  12 & $\leq 0.01$  & 41.1 $\pm$ 0.2 & $\geq 5000$ \\
      & C IV         & 1548.2 &  7415.7 $\pm$ 0.37 & 3.7899 $\pm$ 0.0004 &  81.4 & 6$\times$900    & \\
      & C IV         & 1550.8 &  7428.1  $\pm$ 0.37 & 3.7899 $\pm$ 0.0004 & \\
      & He II        & 1640.5 &  7855.7 $\pm$ 1.07 & 3.7887 $\pm$ 0.0007 & 73.9 $\pm$ 2.7 & 2820 $\pm$ 105 & $\leq 0.01$ &  9.5 $\pm$ 0.3 &  $\geq 790$ \\
      & C III]       & 1908.7 &  9140.7 $\pm$ 1.70 & 3.7889 $\pm$ 0.0007 & \\

\squ\ (total) 
      & \lya         & 1215.7 & 5824.59 $\pm$ 0.18 & 3.7913 $\pm$ 0.0002 & 44.9 $\pm$ 0.5 & 2310 $\pm$  25 & $\leq 0.01$ & 56.1 $\pm$ 0.7 & $\geq 4000$ \\
      & C IV(doublet)& 1549.0 & 7420.75 $\pm$ 1.34 & 3.7907 $\pm$ 0.0009 & 36.5 $\pm$ 3.3 & 1475 $\pm$ 130 & $\leq 0.01$ &  5.0 $\pm$ 0.3 & $\geq 600$ \\
      & He II        & 1640.5 & 7848.50 $\pm$ 4.41 & 3.7843 $\pm$ 0.0027 & 84   & 3220 $\pm$ 4400 &  $\leq 0.02$ &     4.0 $\pm$   1.6 &  $\geq 230$ \\
      & C III]       & 1908.7 & 9141.88 $\pm$ 1.82 & 3.7896 $\pm$ 0.0010 & 59.9 $\pm$ 4.7 & 1960 $\pm$ 155 &  $\leq 0.01$ &  0.2 $\pm$ 0.8 & $\geq 14$  \\
\squ\ (filament) 
      & \lya         & 1215.7 & 5818.07 $\pm$ 0.29 & 3.7859 $\pm$ 0.0002 & 20.0 $\pm$ 0.7 & 1030 $\pm$  36 &  $\leq 0.01$ & 6.2 $\pm$ 0.2 & $\geq 6600$  \\
\squ\ (nearby galaxy) 
      & [O II]       & 3727.0 & 6973.30 $\pm$ 0.33 & 0.8710 $\pm$ 0.0001 & 30.0 $\pm$ 0.8 & 1290 $\pm$  34 &  $\leq 0.06$ & 9.0 $\pm$ 0.1 & $\geq 140$ \\
\squ\ (galaxy at filament) 
      & [O II]       & 3727.0 & 7046.98 $\pm$ 4.25 & 0.8908 $\pm$ 0.0011 & 51.7 $\pm$ ? & 2200 $\pm$ 450 &  $\leq 0.02$ &  3.7 $\pm$  0.2 & $\geq 180$\\

\tur\ (core)  
      & \lya         & 1215.7 & 5834.5  $\pm$ 0.1  & 3.8002 $\pm$ 0.0001 &    & 613/1373 $\pm$ 13/45 &   &   146 $\pm$ 5.5 &   \\
      & C IV         & 1548.2 & 7428.9  $\pm$ 0.2  & 3.7984 $\pm$ 0.0001 &    & 541    $\pm$   14    &   &   7.5 $\pm$  0.2 &    \\
      & C IV         & 1550.8 & 7441.2  $\pm$ 0.2  & 3.7984 $\pm$ 0.0001 &    & 541    $\pm$   14    &   &   5.7 $\pm$  0.2 &    \\
      & HeII         & 1640.5 & 7870.7  $\pm$ 0.4  & 3.7979 $\pm$ 0.0002 &    & 553    $\pm$   28    &   &   5.5 $\pm$  0.3 &    \\
      & C III]       & 1908.7 & 9152.5  $\pm$ 0.6  & 3.7951 $\pm$ 0.0003 &    & 511/1120 $\pm$ 151/135 & &   9.1 $\pm$  0.2 &    \\ 
      & [O II]       & 3727.0 & 17890.36 $\pm$ 0.54 & 3.8002 $\pm$ 0.0001 & 42.4 $\pm$ 1.1 &  711 $\pm$  18 & 57.8 $\pm$  7.1 &   &   \\
      & \hbeta       & 4861.0 & 23351.72 $\pm$ 3.27 & 3.8039 $\pm$ 0.0007 & 33.0 $\pm$ 8.1 &  424$\pm$ 104 & 0.19 $\pm$ 0.04 & 17.2 $\pm$ 4.9 & 92.7 $\pm$ 32.9 \\
      & [O III]      & 4959   & 23816.29 $\pm$ 1.20 & 3.8026 $\pm$ 0.0002 & 36.1 $\pm$ 3.0 &  454 $\pm$  38 & 0.18 $\pm$ 0.04 & 59.8 $\pm$ 5.2 & 337.8 $\pm$ 81.2 \\
      & [O III]      & 5006.9 & 24046.74 $\pm$ 0.42 & 3.8027 $\pm$ 0.0001 & 37.9 $\pm$ 1.1 &  472 $\pm$  13 & 0.18 $\pm$ 0.04 & 222.8 $\pm$ 5.4 & 1260 $\pm$ 295 \\ 

\tur (total) 
     & \hbeta        & 4861   &  23355.84 $\pm$ 8.35 & 3.8047 $\pm$ 0.0017 & 55  &  700 $\pm$ 270 &   0.47$\pm$0.12 & 52.5 $\pm$ 21.5 & 112 $\pm$ 54 \\
     & [O III]       & 4959   &  23810.62 $\pm$ 3.03 & 3.8015 $\pm$ 0.0006 & 50.8 $\pm$ 7.8 &  640$\pm$98 & 0.50$\pm$0.14 &  129.4$\pm$20.5 &  259$\pm$85 \\
     & [O III]       & 5006.9 &  24041.56 $\pm$ 1.13 & 3.8017 $\pm$ 0.0002 & 58.4 $\pm$ 3.0 &  728 $\pm$  37 & 0.47$\pm$0.16 & 484.7$\pm$22.6 & 1031$\pm$351 \\
\enddata
\normalsize
\tablecomments{Where two values are given, line is split.}
\end{deluxetable}

\end{document}